%% file: FinalPaper.tex
\documentclass[10pt,journal]{IEEEtran}
\ifCLASSOPTIONcompsoc
\else
\fi

\ifCLASSINFOpdf
\else
\fi

\usepackage{cite}      
\usepackage{graphicx}  
    
\usepackage{amssymb}
\usepackage{mathrsfs}
\usepackage{amsmath}   

\usepackage[sans]{dsfont}
\usepackage{stmaryrd}
\usepackage{pifont}
\usepackage{wasysym}
\usepackage{algorithm,algorithmicx,algpseudocode}
\usepackage{array}
\usepackage{url}
\usepackage[usenames]{color}
\usepackage[tight,footnotesize]{subfigure}
\usepackage{multirow}

\newcommand{\subparagraph}{}
\usepackage{titlesec}
\titlespacing*{\subsection}{0pt}{0.5\baselineskip}{0.2\baselineskip}

\newtheorem{assumption}{Assumption}

\newtheorem{theorem}{Theorem}

\newcommand{\raf}[1]{(\ref{#1})}

\algnewcommand{\LeftComment}[1]{\Statex \(\triangleright\) #1}

\begin{document}
%
\title{Efficient 3D Road Map Data Exchange for Intelligent Vehicles in Vehicular Fog Networks}
%
%
%
%

\author{Ivan Wang-Hei Ho, \textit{Senior Member, IEEE}, Sid Chi-Kin Chau, \textit{Member, IEEE}, Elmer R. Magsino, Kanghao Jia
\IEEEcompsocitemizethanks{\IEEEcompsocthanksitem Manuscript received May 24, 2019; revised August 23, 2019; accepted December 23, 2019.  This work was supported in part by the General Research Fund (Project No. 15201118) established under the University Grant Committee (UGC) of the Hong Kong Special Administrative Region (HKSAR), China; and by The Hong Kong Polytechnic University (Projects G-YBR2, G-YBXJ).
}
\IEEEcompsocitemizethanks{\IEEEcompsocthanksitem I. W.-H. Ho is with the Department of Electronic and Information Engineering, The Hong Kong Polytechnic University (email:ivanwh.ho@polyu.edu.hk).  
}
\IEEEcompsocitemizethanks{\IEEEcompsocthanksitem S. C.-K. Chau is with the Australian National University (email: sid.chau@anu.edu.au). 
}
\IEEEcompsocitemizethanks{\IEEEcompsocthanksitem E. R. Magsino is with the Department of Electronic and Information Engineering, The Hong Kong Polytechnic University and also with De La Salle University, Manila, Philippines (email: 
elmer.magsino@connect.polyu.hk).   
}
\IEEEcompsocitemizethanks{\IEEEcompsocthanksitem K. Jia is currently with ASTRI, Hong Kong (email: khjia@astri.org). 
}
\IEEEcompsocitemizethanks{\IEEEcompsocthanksitem This paper appears in IEEE Transactions on Vehicular Technology. DOI:0.1109/TVT.2019.2963346. 
}
}%

\markboth{}%
{...}
%


\maketitle

\begin{abstract}
\input{abstract}

\end{abstract}



%
\IEEEpeerreviewmaketitle

\input{Intro}

\input{SysModel}

\input{FormDef}
\input{Index}

\input{Download}

\input{Upload}

\input{Testbed}

\input{Sim}

\input{Conclusion}

\bibliographystyle{IEEEtran}
\bibliography{paperbib}

\end{document}

%% file: abstract.tex
Through connecting intelligent vehicles as well as the roadside infrastructure, the perception range of vehicles can be significantly extended, and hidden objects at blind spots can be efficiently detected and avoided. To realize this, accurate road map data must be downloaded in real time to these intelligent vehicles for navigation and localization purposes. Besides, the cloud must be updated with dynamic changes that happened in the road network. These involve the transmissions of high-definition 3D road map data for accurately representing the physical environments. In this work, we propose solutions under the fog computing architecture in a heterogeneous vehicular network to optimize data exchange among intelligent vehicles, the roadside infrastructure, as well as regional databases. Specifically, the efficiency of 3D road map data dissemination at roadside fog nodes is achieved by exploiting index coding techniques to reduce the overall data load, while opportunistic scheduling of heterogeneous transmissions can be done to judiciously manage network resources and minimize operating cost. In addition, 3D point cloud coding and hashing techniques are applied to expedite the updates of various dynamic changes in the network. We empirically evaluate the proposed solutions based on real-world mobility traces of vehicles and 3D LIght Detection And Ranging (LIDAR) data of city streets. The proposed system is also implemented in a multi-robotic testbed for practical evaluation.

%% file: Intro.tex
\section{Introduction}

\IEEEPARstart{T}{here} have been a plenty of flourishing developments for intelligent vehicles in the past decade. In 2011, Google introduced a driverless car that is tested in real-world streets \cite{Hard2015}. Since then, there have been many self-driving car projects (e.g., Uber \cite{Uber2018}, Waymo \cite{Waymo2019}) gearing towards full driving autonomy.  
Intelligent vehicles are equipped with a plethora of on-board sensors for sensing {the surrounding} environment, and a communication system capable of short-range broadcast and cellular {communications} for information sharing among intelligent vehicles and infrastructure {nodes,} such as roadside units (RSUs), base stations, local controllers, databases, and cloud servers.  {Infrastructure nodes} are fixed-location systems that transmit and receive short- and long-range communications from vehicles for storage, processing, and information exchange.  Collectively, intelligent vehicles and the infrastructure are the fundamental building blocks of a vehicular network.

\begin{figure}[htb!] 
	\centering 
	\includegraphics[scale=1.0]{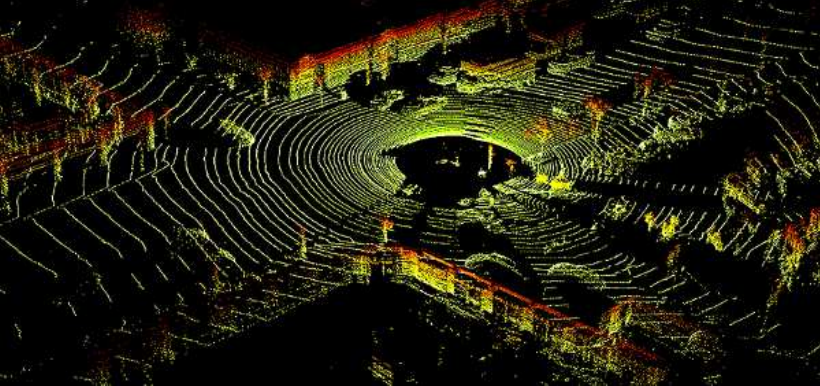} 
	\caption{\small A four-way junction 3D point cloud captured by a LIDAR.}\label{fig:pointcloud}
\end{figure}

In a vehicular network comprising of multiple intelligent vehicles and infrastructure {nodes}, sharing local surrounding information enables the delivery of various vehicular applications and services for improving road safety and travel convenience \cite{CMHC2016VTC}. To facilitate autonomous driving, accurate road map data depicting real-time road events are crucial and should be exchanged among intelligent vehicles and {the} infrastructure for driving perception, localization, route planning, and control.  The road map data capture the static (e.g., buildings, road structures) and dynamic (e.g., presence of road accidents, traffic conditions) features of the road setup.  A particular type of data that can accurately describe the road environment is the {\em 3D LIDAR point cloud data} \cite{kim2011urban}.  An example is illustrated in Fig. \ref{fig:pointcloud} \cite{ford}. This is a set of data points in a 3D coordinate system that represent the surfaces of physical objects in {the} 3D space. However, 3D LIDAR point cloud data {are usually huge in size}. Commercial LIDAR with 64 laser sensors can generate up to 2.2M points per second for the 3D representation of its surrounding environment \cite{velodynelidar}. Given this, the exchange of 3D map data from one vehicular node to another is a challenging task.  Overcoming the bandwidth limitation in highly-dynamic vehicular networks for exchanging 3D point cloud data can enable collaborative perception among vehicular nodes for extending their sights to reach hidden and distant on-road objects or pedestrians.   

\begin{figure}[htb!] 
	\centering 
	\includegraphics[scale=1.2]{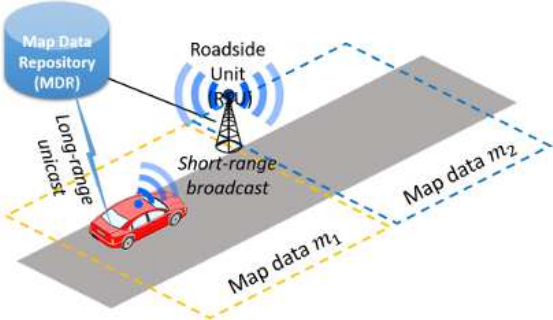}
	\caption{\small Road map information exchange among an intelligent vehicle, a roadside unit, and the map data repository in a V2X network.}\label{fig:parties}
\end{figure}

The information exchanges {among} intelligent vehicles {and the} roadside infrastructure, as illustrated in Fig. \ref{fig:parties}, are supported by Vehicle-to-Everything (V2X) communications (e.g., Vehicle-to-Vehicle (V2V)  and Vehicle-to-Infrastructure (V2I)), which can be realized by either short-range local broadcast or long-range unicast via the cellular network.  In Fig. \ref{fig:parties}, the Map Data Repository (MDR) is another roadside infrastructure that functions as a central database for all map data. It is a cloud computing server that stores the global view of map data over time, merges multiple map data sources, and extracts useful information to assist decision-making at individual vehicles. 

Short-range local broadcast can be achieved by Dedicated Short-Range Communications (DSRC) \cite{kenney2011dedicated} {or 3GPP Cellular V2X (C-V2X) \cite{5g2016case}}, which has been proposed to broadcast basic safety messages (e.g., speed, heading, and location).  However, local broadcast suffers from the limited available spectrum and restricted data transmission rate.  On the other hand, long-range unicast via LTE may be inefficient to share common data among nearby transmitters, such as map data for vehicles in the vicinity. In addition, cellular networks incur {service charges} by mobile service operators.  

Given the increasing number of vehicles on road, estimated over 1.3 billion worldwide in 2016 \cite{CarPop}, {there is an} abundant source of road map information {available, and hence the uploading and downloading of road map data among vehicles and the road infrastructure is time-consuming and takes up a huge amount of network resources}. {In addition, many V2X applications are time-critical and failure in transmission may} lead to accidents  {and casualties} \cite{TeslaAcc}.  {Therefore, how} to effectively manage the information exchange among vehicles and the infrastructure in a heterogeneous V2X network is a pivotal challenge.  In such case, a new computing paradigm is needed to reduce latency in data {processing and communications} so that vehicles {and the} infrastructure can {acquire the required data on time for} making real-time on-road decisions.

\begin{figure*}[htb!] 
	\centering 
	\includegraphics[scale=0.63]{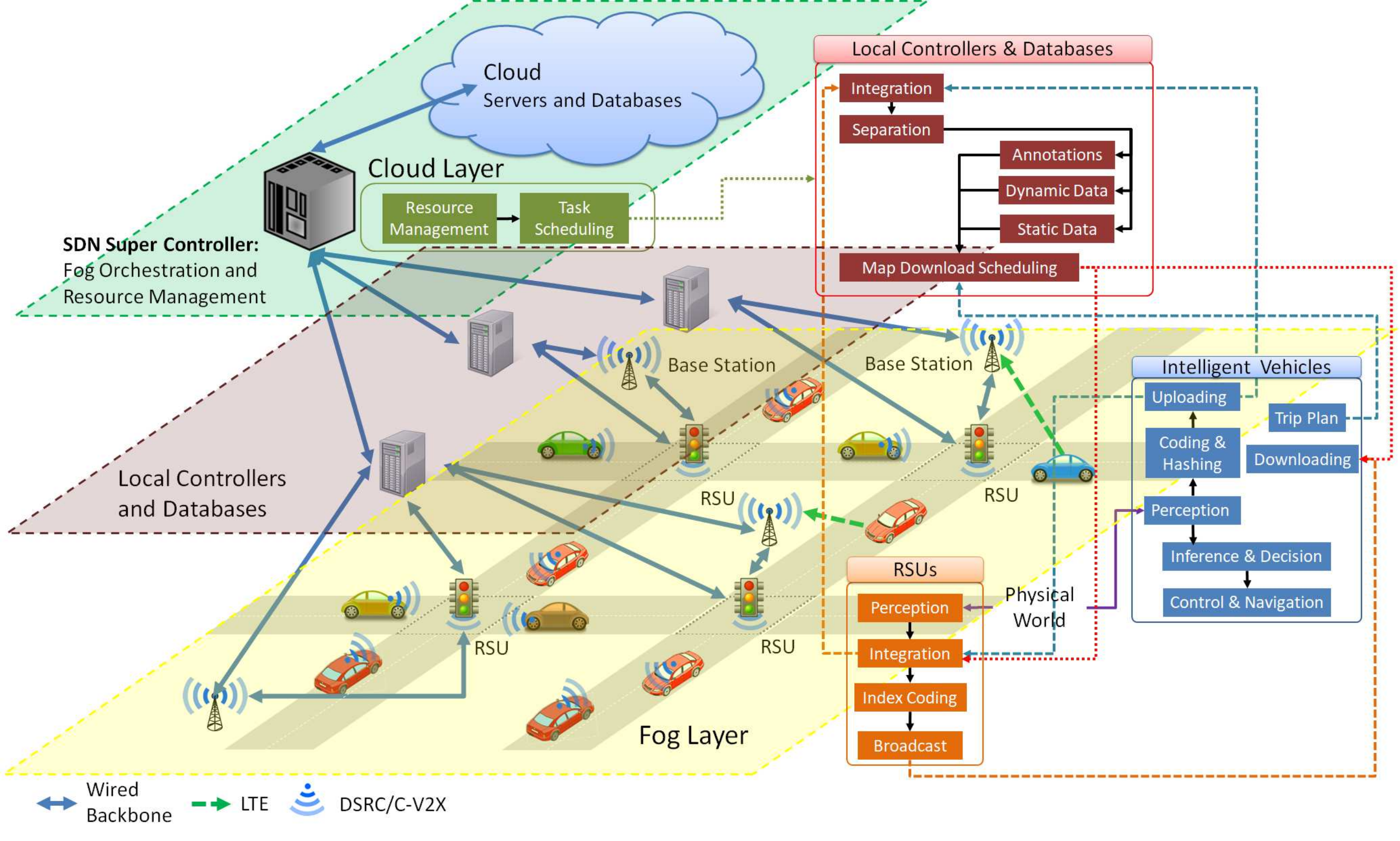}
	\caption{\small {The vehicular fog computing architecture.}  Most information exchange {and computation take} place in the fog layer.}\label{fig:VFCarchi}
\end{figure*}

Fog computing, first coined and introduced by Cisco Systems in 2012, is a recent paradigm bringing cloud computing closer to the network {edges to reduce the latency in various real-time services} \cite{mahmud2018fog}.  The incorporation of fog computing into vehicular networks establishes the Internet of Vehicles (IoV) concept or the vehicular fog computing (VFC) paradigm.  Extensive fog computing surveys \cite{mouradian2017comprehensive,perera2017fog,huang2017vehicular} have outlined the possible application of such computing paradigm in vehicular networks.  {In vehicular fog networks, {as depicted in Fig. \ref{fig:VFCarchi}}, intelligent vehicles act as sensing devices that gather and pre-process surrounding data  {before uploading.  Some data coding and hashing techniques {can also be done at the network edges to} alleviate the traffic load as well.  Intelligent vehicles may {serve as} mobile fog nodes for implementing localized computational tasks and can directly communicate with nearby vehicles via DSRC/C-V2X, especially {when} vehicles are {beyond} the infrastructure's coverage.}  On the other hand, infrastructure {nodes}, such as roadside units (RSUs), traffic lights, base stations, can act as fog nodes for efficiently communicating with intelligent vehicles within its transmission range.  These infrastructure {nodes can} also store huge amount of data and perform computationally-intensive processing and calculations instead of allowing the cloud to do it, {thereby, providing real-time and reliable vehicular applications, {e.g.,} autonomous driving in a dynamic environment.}   
	
{Meanwhile, the data exchange among closely related infrastructure fog nodes in a local region is facilitated by a local controller. Local controllers determine the {transmission mode}, i.e., either long-range unicast (LTE) or short-range broadcast via DSRC or C-V2X.  {They} also decide whether certain road map data are to be forwarded to the map data repository (cloud) \cite{liang2017integrated} or stored in {local} databases.}  {The access of road information from other local regions is administrated by the super} software-defined networking (SDN) controller{, which is the network component with global intelligence \cite{wen2017fog} that} {orchestrates} data traffic and {manages} resources among local controllers and databases \cite{tomovic2017software}.}  {The SDN controllers also perform scheduling of tasks among fog nodes.}  Finally, the map data repository is a cloud {node with} global knowledge of an urban {area} for monitoring and control in a city-wide level \cite{huang2017vehicular}.

Overall, in a vehicular network employing the fog computing paradigm, we consider a scheduled information dissemination mechanism utilizing index coding, data hashing and heterogeneous transmission options. The {major contributions} of this {work} are summarized as follows.
\begin{enumerate}
	\item Under {the vehicular fog computing} framework, we {integrate the index coding algorithm to optimally disseminate high-definition 3D road map data {among} intelligent vehicles and {the} roadside infrastructure  to reduce the number of {required} transmissions and data load while satisfying {the vehicular} demands.}
	\item {We propose fog-based opportunistic scheduling algorithms based on vehicular trip plans {for map data downloading} in city-wide vehicular networks. These dynamic schedulers determine the mode of transmission (short-range broadcast or long-range unicast) based on the available resources at fog devices to reduce the overall operating cost of the network. In addition, differential coding and hashing techniques for 3D point cloud data uploading at the vehicular level is also proposed to avoid data redundancy, and hence reduce the processing and computation load of roadside fog nodes.}
	\item Utilizing empirical mobility traces and 3D LIDAR data of city streets, {we rigorously evaluate the performance of the proposed algorithms and system.}  We have also implemented our system in a multi-robotic vehicle testbed for practical evaluation.
\end{enumerate}

The paper is organized as follows. Section II {describes our proposed information dissemination system at the fog layer}. In Section III, {we formally define the information dissemination problem and discuss the downloading and uploading operations of 3D road map data.}  Section IV {presents motivating examples on {utilizing} index coding {for vehicular data exchange, and derives the optimal index coding scheme for single road junction as well as the city-wide multi-junction scenario.}  {The fog-based opportunistic scheduling problem is tackled in Section V, and}  {the techniques for efficient uploading of 3D LIDAR point cloud data from vehicles is covered in Section VI.}  {Sections VII to VIII {present experimental and simulation results obtained based on our multi-robotic vehicle testbed and empirical mobility traces, respectively.}  Finally, Section IX concludes this paper.

%% file: SysModel.tex
\section{Information Exchange of 3D Road Map Data in V2X Networks}

{To implement efficient {road map data dissemination in a} vehicular fog network, we propose the 3D MAp Dissemination System ({3D-MADS}).}  The general operation of {3D-MADS} includes intelligent vehicles, roadside units, local controllers and databases, which are all within the fog layer in Fig. \ref{fig:VFCarchi}.
Overall, the system distributes map data among the parties in a timely manner, taking into account the characteristics of long-range unicast and short-range broadcast transmissions. Short-range broadcast normally has limited spectrum resources at lower transmission cost, while long-range unicast has large bandwidth capacity at higher transmission cost. We aim at optimizing these transmission options while satisfying the dynamic data demand of respectively vehicles. By referring to Fig. \ref{fig:VFCarchi}, each component or module in 3D-MADS and its corresponding tasks are explained as follows.

\begin{itemize}
	\item {\bf Intelligent Vehicles}
	
	- {\em Uploading} enables sharing of on-board LIDAR data among vehicles via the vehicular fog network.
	
	- {\em Coding \& Hashing} encodes and identifies differentiated data and redundant map information.
	
	- {\em Downloading} delivers the most updated 3D road map data from local databases to intelligent vehicles via either cellular network or local broadcast at RSU fog nodes. 
	
	- {\em Perception} utilizes on-board sensors, e.g., LIDAR and GPS, to perceive the surrounding road environment as 3D point cloud data, from which we can detect and recognize objects and obstacles in the environment. The locally processed 3D point cloud data will be uploaded to RSUs or local controllers for further integration with the data from other vehicular and roadside nodes.  
	
	- {\em Inference \& Decision} allows intelligent vehicles to predict their movements for autonomous navigation and control based on the perceived and downloaded 3D road map data as well as position information.
	
	- {\em Control \& Navigation} relies on driving feedback and manages the intelligent vehicles to move safely and appropriately in the environment.
	
	\smallskip
	\item {\bf Roadside Unit (RSU) Fog Nodes}

	- {\em Perception} provides blind-spot views that cannot be detected by intelligent vehicles via the local sensors.
	
	- {\em Integration} combines downloaded 3D road map data from the cloud with the local LIDAR sensor data before sending them to nearby intelligent vehicles.
	
	- {{\em Index Coding} encodes 3D road map data according to the data demand and availability of nearby vehicles to improve the transmission efficiency.}
	
	- {\em Broadcast} is the periodic transmission of {index-coded} data to nearby vehicles via local short-range broadcast. 
	
	\bigskip
	\item {\bf Local Controllers and Databases (LCD)}
	
	- {\em Integration} coordinates the data exchanged among intelligent vehicles and RSU by setting the locations and boundaries of each region of map data. It can also correct and realign the LIDAR data from different vehicles that may contain drifting inaccuracy.
	
	- {\em Separation} differentiates static and dynamic objects in the integrated 3D road map data via segmentation. Additional annotations can be generated based on machine learning techniques \cite{wang2019vtc} to label the objects in the map data. Different coding and transmission schemes can be applied to data with different characteristics.
	
	- {\em Scheduling} organizes the download and upload transmissions based on the trip plans of vehicles, given the options of using either the cellular network unicast or the short-range local broadcast transmissions.
	
\end{itemize}

With respect to Fig. \ref{fig:VFCarchi}, we can see that 3D-MADS is an interdisciplinary system that requires the joint effort from multiple fields (e.g., communications, signal processing, computing, navigation and control, transportation engineering, etc.), which is our long-term goal. In this paper, we focus on investigating and discussing data exchange related modules (which include index coding, map download scheduling, and coding and hashing) to kick start the development of such system.

%% file: FormDef.tex
\section{Formulation and Definitions} \label{sec:model}

In this section, we formally define the 3D road map data dissemination problem for intelligent vehicles. Consider a set of discrete time slots $t \in {\cal T}$, where $|{\cal T}|=T$, and a network of roads that is represented by graph ${\cal G}=({\cal N}, {\cal E})$, where each node $v \in {\cal N}$ represents a junction and each undirected edge $e \in {\cal E}$ represents a road segment. For each edge $e$ at time $t$, a set of map data is associated and denoted by $m_e(t)$. $m_e(t)$ consists of both static data set $m^{\sf s}_e $ and dynamic data set $m^{\sf d}_e(t)$, such that $m_e(t) = m^{\sf s}_e \cup m^{\sf d}_e(t)$. We consider an abstract representation, without specifying the elements in $m_e(t)$. That is, one may consider an element in $m_e(t)$ as a map data file. The dynamic data may be generated from roadside sensors, and perception from other vehicles.  For practicality, we consider the dynamic data within a certain time window $\tau$ from the current time $t$, namely $m^{\sf d}_e(t') $ where $t' \in [t- \tau, t]$.

There is a set of vehicles ${\cal C}$ where each vehicle $c \in {\cal C}$ is associated with a trip plan ${\sf P}^{c}$, which is a path in ${\cal G}$. We represent ${\sf P}^{c}$ by a set of edges in ${\cal E}$, or a sequence of nodes in ${\cal N}$. Let the time of vehicle $c$ entering edge (i.e., road segment) $e \in {\sf P}^{c}$ be ${\sf t}^{c}_{e}$, and the time of entering node (i.e., junction) $v \in {\sf P}^{c}$ be ${\sf t}^{c}_{v}$.

\subsection{Downloading} 

 Each vehicle $c \in {\cal C}$ downloads both static data $m^{\sf s}_e $ before ${\sf t}^{c}_{e}$, and dynamic data $m^{\sf d}_e(t')$, for some $t' \in [{\sf t}^{c}_{e} - \tau, {\sf t}^{c}_{e}]$, at some time between $t'$ and ${\sf t}^{c}_{e}$. The options for downloading are either using short-range broadcast transmissions at RSUs, or unicast transmissions via cellular networks. We assume that LTE cellular network transmissions have much larger capacity, whereas short-range broadcast transmissions are limited by local spectrum allocation. On the other hand, the short-range broadcast transmissions incur no or very low costs, whereas cellular network  transmissions incur higher costs. 

 We assume that RSU fog nodes are only located at a subset of nodes in ${\cal G}$, denoted by ${\cal R} \subseteq {\cal N}$. Denote the set of edges connecting to RSU fog node $r \in {\cal R}$ by ${\cal E}_r \subseteq {\cal E}$.  A vehicle $c$ can receive data from $r$, when entering edge $e \in {\cal E}_r$. At each RSU fog node $r \in {\cal R}$, there is a download capacity of ${\rm C}^{\downarrow}_r$ at $r$, whereas there is no capacity limit via cellular networks. 
 
 Let the data transmitted by RSU fog node $r$ using short-range broadcast at time $t$ be $x_r(t)$. A vehicle $c$ can also download data via cellular networks, which is denoted by $y^{c}(t)$. At the time $t$, let $X^{c}(t)$ be the union of all data  that $c$ has received from the visited RSU fog nodes on its path before time $t$, namely,
\begin{equation}
X^{c}(t) \triangleq \bigcup _{e \in  {\sf P}^{c}  \wedge e \in {\cal E}_r \wedge {\sf t}^{c}_{e} \le t} \Big\{x_r({\sf t}^{c}_{e})\Big\}
\end{equation}

Also, let $Y^{c}(t)$ be the union of data that $c$ has received from cellular network transmissions before time $t$, namely,

\begin{equation}
Y^{c}(t) \triangleq \bigcup _{t'\le t } \Big\{y^{c}(t')\Big\}
\end{equation}

We denote a decoding function by ${\sf Dec}[\cdot]$, which decodes all the downloaded data to a set of map data, $M^{c}(t) = {\sf Dec}[X^{c}(t), Y^{c}(t)]$.

We aim to minimize the number of cellular network transmissions, subject to the constraints of timely delivery of static and dynamic data:
\begin{align}
& \min_{\{x_r(t), y^{c}(t)\mid   t \in {\cal T},  c \in {\cal C}, r \in {\cal R} \}}   \sum_{c \in {\cal C}}\big|Y^{c}(T)\big|\\
& \mbox{subject to\ }  |x_r(t)| \le  {\rm C}^{\downarrow}_r,\ \mbox{for all }   t \in {\cal T}, r \in {\cal R}, \label{cons:dlcapacity} \\
& \ \qquad \qquad m^{\sf s}_e \in M^{c}({\sf t}^{c}_{e}), \ \mbox{for all }  c \in {\cal C}, e \in {\sf P}^{c}, \label{cons:stdata}\\
& \ \qquad \qquad m^{\sf d}_e(t) \in M^{c}({\sf t}^{c}_{e}), \ \mbox{for all }  c \in {\cal C}, e \in {\sf P}^{c},  \notag \\
& \ \qquad \qquad \qquad \qquad \qquad \quad \mbox{for some } t \in [{\sf t}^{c}_{e} - \tau, {\sf t}^{c}_{e}]. \label{cons:dydata} 
\end{align}
In this problem, we assume that the trip plans of all vehicles are given a-priori. However, the online version is also discussed in Sec.~\ref{sec:schedule}.
Cons.~\raf{cons:dlcapacity} represents the capacity constraint of local broadcast, whereas Cons.~\raf{cons:stdata} and Cons.~\raf{cons:dydata} represent the download constraint of static data and  dynamic data, respectively.

\subsection{Uploading} 

The previous section considers downloading map data from fog units (e.g., RSUs and base stations). In practice, intelligent vehicles are equipped with various sensors (e.g., LIDAR, RADAR, camera, inertial measurement unit (IMU), GPS unit, etc.), whose data can be uploaded to LCD via RSUs or base stations for sharing with other vehicles. 

We consider the uploading of processed 3D LIDAR point cloud data in this paper, in which the operations can be optimized by uploading hash files of the perception data and differentially coded data to reduce the redundant data load to the network, as described in Sec.~\ref{sec:octree}.

%% file: Index.tex
\section{Index Coding for Local Broadcast at {RSU Fog Nodes}} \label{sec:index}

To reduce latency in the presence of numerous intelligent vehicles, the local broadcast operations at RSU {fog nodes} can be improved by index coding. Index coding is a variant of network coding \cite{ElRouayheb2010c,Bar-Yossef2011} applied to wireless communications. Nearby vehicles will likely receive common information by local broadcast, which also possess certain prior information (i.e., information received from other RSU fog nodes at previously traversed road segments). We show that smart data dissemination considering prior information can significantly reduce the number of broadcast transmissions needed. 

This section only considers the dissemination of static data, without capacity constraint. In the next section, we will develop heuristics for the settings with capacity constraint and dynamic data. It is assumed that the local broadcast transmissions incur a very low cost, which is negligible.

\subsection{Motivating Examples}

We first present some motivating examples of index coding. The basic idea of using index coding to optimize transmissions at RSU fog nodes is by mixing the transmitted packets with prior information previously received. 

\begin{figure}[htb!] 
	\centering 
            \includegraphics[scale=0.6]{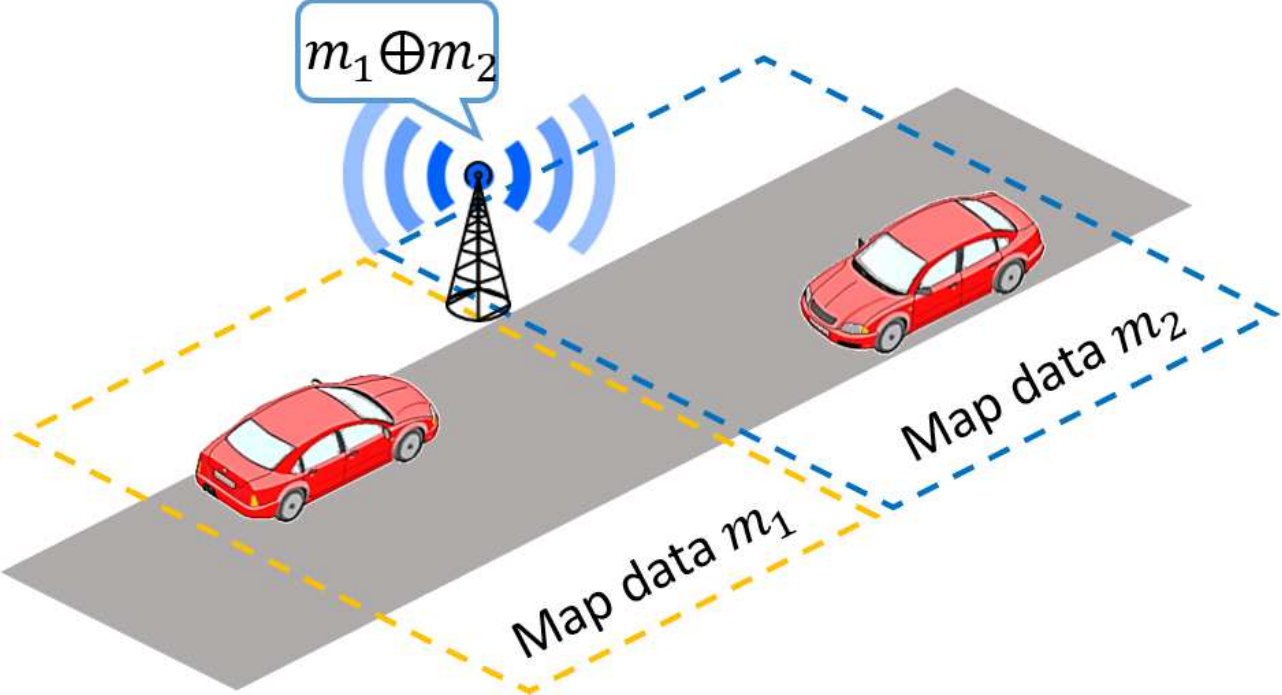}
        \caption{\small An example of index coding for map data dissemination with two opposite traveling intelligent vehicles.}\label{fig:2way}
\end{figure}

{\bf Example 1}:
We illustrate a simple example using index coding for map data dissemination in Fig.~\ref{fig:2way}. There are two intelligent vehicles traveling on opposite directions.  Consider the static map data for two road segments, denoted by $m_1$ and $m_2$ in bit string representation. Both vehicles are now within the transmission range of a common RSU fog node and had obtained map data $m_1$ and $m_2$ correspondingly, before entering their respective road segments. The common RSU fog node can broadcast a coded packet $m_1{\oplus}m_2$, where $\oplus$ is a bitwise XOR operator, thereby, reducing the number of broadcast transmissions. To obtain the required map data, the vehicles can decode using the received data as follows: $m_1{\oplus}(m_1{\oplus}m_2) = m_2$ and $m_2{\oplus}(m_1{\oplus}m_2) = m_1$.  

\begin{figure}[htb!] 
	\centering 
            \includegraphics[scale=1.0]{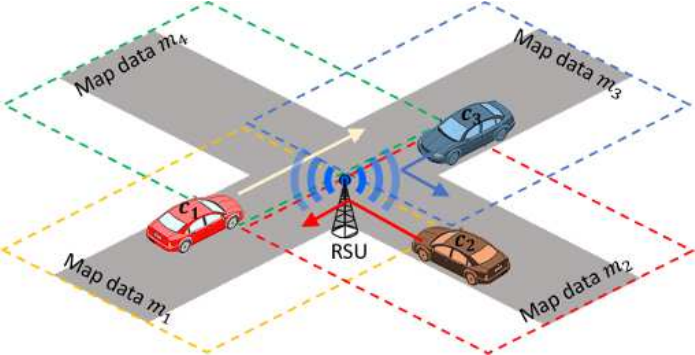}
        \caption{\small A four-way junction with three vehicles: $c_1$ with direction $m_1{\to}m_3$, $c_2$ with direction $m_2{\to}m_1$, and $c_3$ with direction $m_3{\to}m_1$.}\label{fig:junc}
\end{figure}

\smallskip

{\bf Example 2}:
We next consider an example of index coding for a four-way junction in Fig.~\ref{fig:junc}. There are three vehicles: $c_1$ moving from $m_1$ to $m_3$, $c_2$ moving from $m_2$ to $m_1$, and $c_3$ moving from $m_3$ to $m_2$. Note that we use $m_i$ to denote road segment $i$ as well as the map data of road $i$ for notation simplicity here. In this case, the RSU fog node only needs to broadcast two packets: $m_1\oplus m_3$ and $m_2\oplus m_1$. $c_1$ can obtain $m_3 =m_1{\oplus}(m_1{\oplus}m_3)$, $c_2$ can obtain $m_1 =m_2{\oplus}(m_2{\oplus}m_1)$, and $c_3$ can obtain $m_2 =m_3{\oplus}(m_2{\oplus}m_1)\oplus (m_1{\oplus}m_3)$.

\medskip

In the preceding examples, the vehicles are able to decode the required packets by bitwise XOR operation $\oplus$. Note that the bitwise XOR operator is a linear operator over the binary number field. Applying index coding in these scenarios can improve network throughput and reduce latency. {One mild drawback is that it generates overheads in the network.  However, since only binary-coded packets are employed in our scheme, it can still be solved within polynomial time.}  {The reader is referred to Section \ref{OHAnalysis} for the overall delay analysis based on the processing overheads and transmission delay in the proposed index coding scheme.}

\subsection{Optimal Index Coding for Single Junction}

In this section, we derive the general theories for constructing index coding schemes for a road network with a-priori trip plans of vehicles. We only consider linear index coding, i.e., the coding/decoding schemes only rely on bitwise XOR operator. In linear index coding, the encoding/decoding operations can sometimes be interpreted as unions and complemented intersections on a set of packets\footnote{For example, coding by $m_1\oplus m_3$ can be interpreted as union $m_1\cup m_3$, whereas decoding by $(m_1\oplus m_3) \oplus m_1 = m_3$ can be can be interpreted as complemented intersection $(m_1\cup m_3) \cap (m_1\cup m_3 \backslash m_1) = m_3$.}.

In general, a good index coding scheme for multiple junctions is a hard problem, because it is related to the multi-source network coding problem, which is an open problem \cite{ElRouayheb2010c,Bar-Yossef2011}. Instead, we focus on one single junction first, and then extend the single-junction scheme as a heuristic for multiple junctions. In fact, under the assumption of `single meeting' as depicted in the next subsection, this is an optimal solution. Note that we ignore the download capacity in this section, which will be considered in the general schemes in the next section.

\begin{figure}[htb!] 
\centering \vspace{-5pt}
	\includegraphics[scale=0.35]{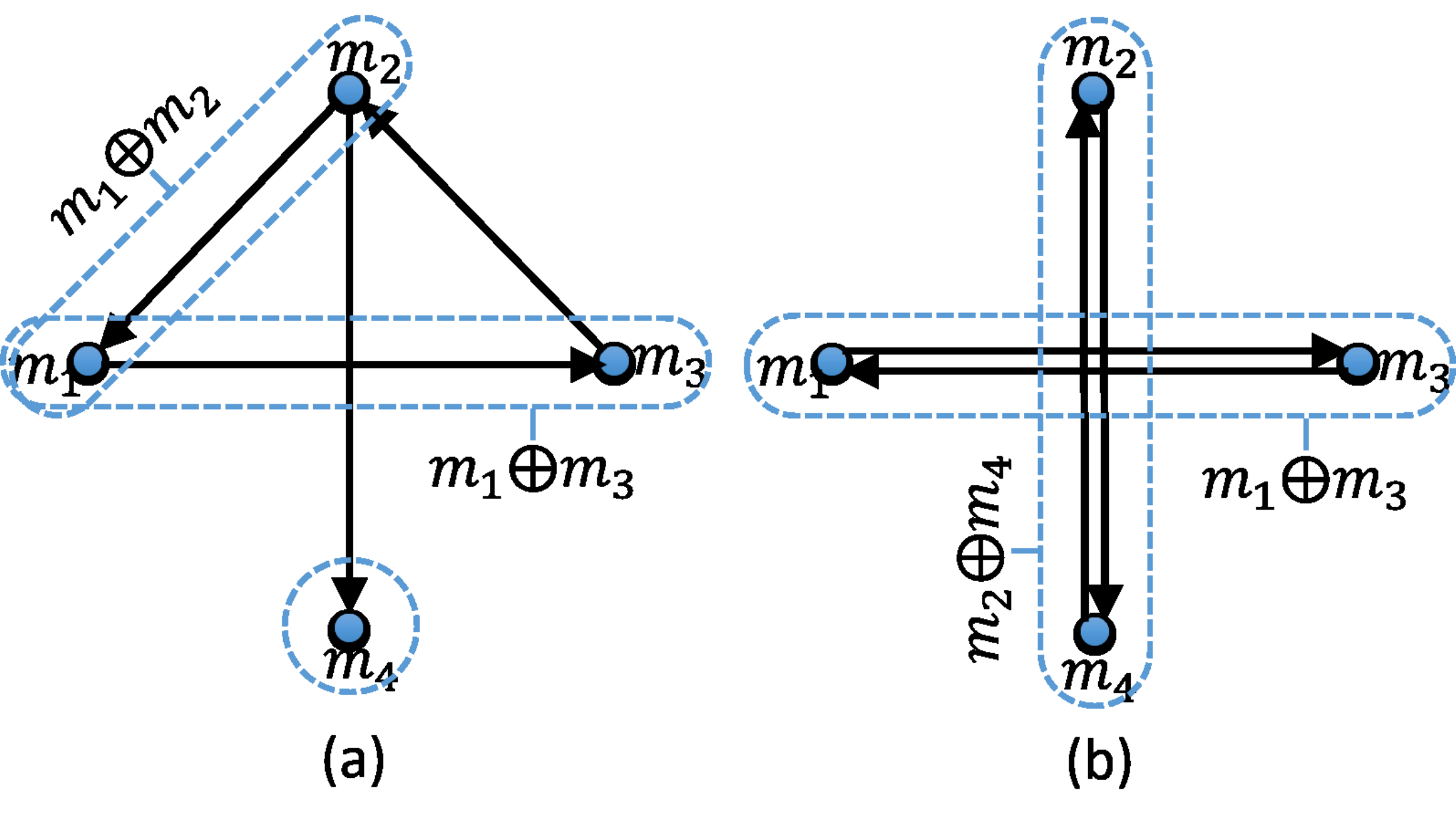}  \vspace{-10pt}
\caption{\small Two examples of demand graph ${\cal D}_r$ and their optimal index coding schemes for a four-way junction. (a) Four vehicles with directions: $(m_1{\to}m_3)$, $(m_3{\to}m_2)$, $(m_2{\to}m_1), (m_2{\to}m_4)$, and an optimal index coding scheme is $\{m_1{\oplus}m_2, m_1{\oplus}m_3, m_4\}$. (b) Four vehicles with directions: $(m_1{\to}m_3)$, $(m_3{\to}m_1)$, $(m_2{\to}m_4)$, $(m_4{\to}m_2)$, and an optimal index coding scheme is $\{m_1{\oplus}m_3, m_2{\oplus}m_4\}$. }\label{fig:4way}
\end{figure}

To construct a good index coding scheme, we consider a particular RSU fog node at a single $n$-way junction, labeled as $r \in {\cal R}$. We represent the demands for map data by a directed graph (called {\em demand graph}) ${\cal D}_r$ with a set of $n$ nodes representing the set of connected road segments to $r$. Denote the map data for the ${k}$-th road segment by $m_{k}$, where $k \in \{1,...,n\}$. There is a directed edge $(m_{k_1}{\to}m_{k_2})$ in ${\cal D}_r$, if there is a vehicle moving from the ${k_1}$-th road segment to the ${k_2}$-th road segment, which needs to obtain $m_{k_2}$, given $m_{k_1}$ as prior information. The destination nodes in ${\cal D}_r$ (i.e., those with at least one in-coming directed edge) are called the {\em demanded packets}. Two examples of ${\cal D}_r$ for a four-way junction are shown in Fig.~\ref{fig:4way}.

The uncoded packets $\{m_1,...,m_n\}$ are called {\em source packets}. A packet consists of $K$ source packets combined by bitwise XOR operator is called a {\em $K$-ary coded packets}. For example, $m_2{\oplus}m_1$ is a binary-coded packet. An index coding scheme, denoted by ${\cal I}$, is a set of coded or source packets. For convenience of analysis, we assume that each packet of map data has a uniform size. If packets have different sizes, padding will be used. 

Given a set of demanded packets, we aim to construct an optimal index coding scheme using the  minimal number of transmitted packets that can be decoded into the required information (i.e., destination nodes in ${\cal D}_r$).  Note that the construction of a decodable index coding scheme is similar to a generalization of the set cover problem. Each coded packet is a cover, while the demanded packets are items to be covered by some coded packets. The decodability of coded packets requires that a combination of complemented intersections (i.e., XOR operations) of the received coded packets can generate the demanded packets.

\medskip

\begin{theorem} \label{thm:2-coded}
Given the demand graph ${\cal D}_r$, an optimal index coding scheme ${\cal I}$ can be constructed using source packets and binary-coded packets. In particular, each demand $(m_{k_1}{\to}m_{k_2})$ in ${\cal D}_r$ can be decoded by one of the following ways:
\begin{enumerate}

\item A source packet (i.e., $m_{k_2} \in {\cal I}$).

\item Or a sequence of connected binary coded packets, say $\{m_{k_1}{\oplus}m_{k_r}, m_{k_r}{\oplus}m_{k_{r-1}}, ..., m_{k_3}{\oplus}m_{k_2}\} \subseteq {\cal I}$, such that the required packet can be decoded by $m_{k_1}$ and such a sequence of binary-coded packets.

\end{enumerate}
\end{theorem}

See two examples of optimal index coding schemes in Fig.~\ref{fig:4way}, where an arrow represents a demand, and a dashed enclosure represents a coded or source packet.

By Theorem~\ref{thm:2-coded}, it suffices to consider binary-coded packets. We next present a polynomial-time algorithm {\sf 1J-IdxCd} to identify the optimal index coding scheme, which first adds any demand $(m_{k_1}{\to}m_{k_2})$ as a coded packet, and then removes redundant packets in any cycles of coded packets, while ensuring the decodability of demanded packets.

\begin{algorithm}
\caption{\small {\sf 1J-IdxCd}$[{\cal D}_r]$}
{\scriptsize
\begin{algorithmic}[1] 
\State ${\cal I} \leftarrow \varnothing$
\For {$(m_{k_1}{\to}m_{k_2}) \in {\cal D}_r$}
\State ${\cal I} \leftarrow {\cal I} \cup \{ m_{k_1}{\oplus}m_{k_2} \}$
\LeftComment{{\em Flag ${\sf lock}_{{k_1},{k_2}}$ prevents $m_{k_1}{\oplus}m_{k_2}$ to be removed from ${\cal I}$}}
\State ${\sf lock}({k_1},{k_2}) \leftarrow {\sf False}$ 
\EndFor
\While{there exists cycle $\{m_{k_1}{\oplus}m_{k_2}, m_{k_2}{\oplus}m_{k_3}$ $..., m_{k_r}{\oplus}m_{k_1}\} \subseteq {\cal I}$}
\For {$m_{k_t}{\oplus}m_{k_{t+1}} \in \{m_{k_1}{\oplus}m_{k_2}, m_{k_2}{\oplus}m_{k_3}$ $..., m_{k_r}{\oplus}m_{k_1}\} $}
\If {${\sf lock}({k_t},{k_{t+1}}) = {\sf False}$}
\State ${\cal I} \leftarrow {\cal I} \backslash \{ m_{k_t}{\oplus}m_{k_{t+1}} \}$
\For {$(m_{k'_1}{\to}m_{k'_2}) \in {\cal D}_r$}
\If {there exists no path $\{m_{k'_1}{\oplus}m_{k'_r}, ..., m_{k'_3}{\oplus}m_{k'_2}\}$ \\
    \quad \qquad \qquad $ \subseteq {\cal I} \backslash \{ m_{k'_1}{\oplus}m_{k'_{2}} \}$}
\State ${\sf lock}({k'_1},{k'_2}) \leftarrow {\sf True}$
\EndIf
\EndFor
\EndIf
\EndFor
\EndWhile
\State \Return ${\cal I}$
\end{algorithmic}
}
\label{alg:1JIdxCd}
\end{algorithm}

\medskip

\begin{theorem} \label{thm:1JIdxCdAlgo}
Algorithm {\sf 1J-IdxCd} produces an optimal index coding scheme for a single junction.
\end{theorem}

{Let us apply Algorithm {\sf 1J-IdxCd} to the example in Fig. \ref{fig:4way}(a). The initial index coding scheme is ${\cal I} \leftarrow \{(m_1\oplus m_2), (m_2\oplus m_3),(m_3\oplus m_1),(m_2\oplus m_4)\}$ and their corresponding locks, {\sf lock}($k_1,k_2$) are set to {\sf False}.  These are defined by lines 2--5.  Lines 6--18 remove the redundant and unnecessary coded packets.  Coded packet $(m_2\oplus m_4)$ will be the first to be removed since {\sf lock}(2,4) is {\sf False} and there is no cycle that includes it.  On the other hand, the combination of the coded packets $\{(m_1\oplus m_2), (m_2\oplus m_3),(m_3\oplus m_1)\}$ forms a cycle.  Any one of the three coded packets can be removed since all their corresponding locks are false. For instance, we remove $(m_2\oplus m_3)$, then the locks {\sf lock}(1,2) and {\sf lock}(1,3) become {\sf True}.  Therefore, Algorithm {\sf 1J-IdxCd} returns the final index coding scheme as ${\cal I} \leftarrow \{(m_1\oplus m_2), (m_3\oplus m_1)\}$.}

Note that Algorithm {\sf 1J-IdxCd} produces an index coding scheme with coded packets only. {\sf 1J-IdxCd} can be improved by replacing some coded packets by source packets, because the source packets are immediately decodable. One can replace any binary-coded packet that is only used to produce just one demanded packet by the corresponding source packet, although the size of ${\cal I}$ will remain the same. {For the example in Fig. \ref{fig:4way}(a), we add the source packet $m_4$ to the index coding scheme ${\cal I}$, making the optimal index coding scheme as ${\cal I} = \{m_1 \oplus m_2, m_1 \oplus m_3, m_4\}$.}

The reader is referred to the Appendix in \cite{Ho2019} for the proofs of Theorems 1 and 2.

\subsection{Extension for Multiple Junctions}

We next present an extension for multiple junctions. The basic idea is to adopt {\sf 1J-IdxCd} as a basis for multiple junctions. We assume that all trip plans, $\{{\sf P}^c\}_ {c \in{\cal C}}$, are given a-priori. We define a {\em meeting relation graph} among the vehicles by ${\cal G}_{\rm meet} = ({\cal N}_{\rm meet}, {\cal E}_{\rm meet})$, where the set of nodes ${\cal N}_{\rm meet}$ are subsets of vehicles $( \subseteq {\cal C})$ having intersected trip plans, and the set of directed edges ${\cal E}_{\rm meet}$ are the temporal ordering between meetings with a common vehicle. Namely,
\begin{align}
{\cal N}_{\rm meet} \triangleq & \Big\{ (c_1,...,c_r) \subseteq {\cal C} \mid \exists v \in  \bigcap_{i=1}^r {\sf P}^{c_i} \mbox{\ and\ }  {\sf t}^{c_1}_v =  ...= {\sf t}^{c_r}_v \Big\} \\
{\cal E}_{\rm meet} \triangleq  &\Big\{  (c_1,c_2,...,c_r) \to  (c_1,c'_2,...,c'_s) \in {\cal N}_{\rm meet} \times {\cal N}_{\rm meet} \notag\\  & \qquad \mid \exists v_1 \in \bigcap_{i=1}^r {\sf P}^{c_i} \mbox{\ and\ }    
 \exists v_2 \in \bigcap_{i=1}^s {\sf P}^{c'_i}  \mbox{\ and\ }   {\sf t}^{c_1}_{v_1} < {\sf t}^{c_1}_{v_2}  \Big\} 
\end{align}

If two vehicles meet in their trip plans, then there are two cases: (1) traveling in different directions (e.g., meeting at a junction), or (2) traveling along with each other. Case (1) is utilized in index coding to broadcast mixed information (via bitwise XOR) to vehicles, and then the vehicles decode the mixed information using different prior information they received previously. However, there will be no impact by index coding for case (2).

\medskip

\begin{assumption}({\bf Single Meeting})
We assume that the trip plans of every pair of vehicles intersect at most {\em once}, namely, if they meet and depart, then they will never meet again. In practice, if the autonomous vehicles always follow the shortest paths and employ deterministic tie-breaking for the paths of equal distance, then the meeting with another autonomous vehicles of different source or destination is only at most once. Otherwise, this will contradict to the property of shortest paths. Since they meet at most once, each meeting event can be uniquely identified as a node in ${\cal N}_{\rm meet}$, and ${\cal G}_{\rm meet}$ is also a directed acyclic graph.
\end{assumption}

\medskip

\begin{theorem} \label{thm:multijunc}
If the single meeting assumption (Assumption 1) holds, then applying Algorithm {\sf 1J-IdxCd} independently at each junction will produce an optimal index coding scheme for multiple junctions.
\end{theorem}

\medskip

The reader is referred to the Appendix in \cite{Ho2019} for the proof of Theorem 3. Note that even if the vehicles meet more than once, Theorem~\ref{thm:multijunc} still provides a heuristic to construct a good index coding scheme for multiple junctions with limited meetings among vehicles.

%% file: Download.tex
\section{{Fog-based Opportunistic Scheduling of Heterogeneous V2X Networks}} \label{sec:schedule}

The previous section considered the basic setting with static map data and the absence of capacity constraint, under the single meeting assumption. In this section, we present scheduling schemes that decide the transmission options for both static and dynamic map data  from the LCD to intelligent vehicles, considering capacity constraint at RSU fog nodes. The scheduling schemes heuristically apply Algorithm {\sf 1J-IdxCd} at each junction.

\subsection{Downloading}

First, denote the starting and ending time of vehicle $c$'s trip plan by $t^{\sf s}_c$ and $t^{\sf d}_c$ respectively.
There are two modes of scheduling:
\begin{enumerate}

\item {\bf Offline Mode}: All the trip plans of intelligent vehicles $\{{\sf P}^{c} \}_{c \in {\cal C}}$ are known in advance. 

\item {\bf Online Mode}: Not all trip plans are known. Only the trip plans of intelligent vehicles started at the current time $t_{\sf now}$ or before (i.e., $\{{\sf P}^{c}\mid t^{\sf s}_c \le t_{\sf now} \}_{c \in {\cal C}}$) are known.

\end{enumerate}

As illustrated in Fig. \ref{fig:VFCarchi}, the LCD scheduler decides the download operations of map data to individual vehicles according to their GPS locations and trip plans. The map data are first downloaded via nearby RSUs (via short-range broadcast). In case of insufficient capacity at the RSU fog nodes, cellular network transmissions will be utilized. 

Recall that static map data is denoted by $m^{\sf s}_e$ and dynamic data by $m^{\sf d}_e(t)$ for each road segment $e \in {\cal E}$. $m^{\sf s}_e$ should be downloaded to vehicle $c$ before or at time ${\sf t}^{c}_{e}$, and $m^{\sf d}_e(t)$ should be downloaded to $c$ at some time between ${\sf t}^{c}_{e} - \tau$ and ${\sf t}^{c}_{e}$. For each RSU $r \in {\cal R}$, let ${\cal C}_r(t) = \{c \in {\cal C} \mid t= {\sf t}^{c}_{r}\}$ be the set of vehicles that meet at junction $r$ at time $t$, and ${\cal D}_r(t)$ be the demand graph considering the vehicles in ${\cal C}_r(t)$. 

Let $x^{\sf d}_r(t, e)$ and $x^{\sf s}_r(t, e)$ be the decisions of the scheduler to broadcast static and dynamic data respectively at RSU fog node $r$, for road segment $e$ at time $t$. Similarly, let $y_{\sf d}^c(t, e)$ and  $y_{\sf s}^c(t, e)$ be the decisions of the scheduler to download static and dynamic data via cellular networks to vehicle $c$. Let ${\cal I}^{\sf s}_r(t)$ and ${\cal I}^{\sf d}_r(t)$ be the index coding schemes for static and dynamic data respectively at time $t$. Also, let us denote the single-junction scheme applied to RSU fog node $r$ for static data by ${\sf 1J\mbox{-}IdxCd}^{\sf s}[{\cal D}_r(t)]$. Similarly, ${\sf 1J\mbox{-}IdxCd}^{\sf d}[{\cal D}_r(t)]$ for the scheme applied to dynamic data in $[t - \tau, t]$ for a given time window $\tau$. We denote the size of the code by $|\cdot|$ (e.g., $|m^{\sf s}_e|$ and $|{\sf 1J\mbox{-}IdxCd}^{\sf s}[{\cal D}_r(t)]|$).

By the single meeting assumption, myopic scheduling of static data at the respective junction in an on-demand manner is optimal.  
For dynamic data, the latest information is always more useful. Hence, myopic scheduling is also desirable. However, in the presence of capacity constraint, it may not be possible to schedule all required transmissions in an on-demand manner. In this case, we have to greedily pick a subset of vehicles at each junction to maximize the efficiency of transmissions. Formally, given a demand graph ${\cal D}_r(t) = ({\cal N}[{\cal D}_r(t) ], {\cal E}[{\cal D}_r(t)] )$, we define subgraph ${\cal H}_r = ({\cal N}, {\cal E})$, where ${\cal N}$ is a subset of ${\cal N}[{\cal D}_r(t)]$ and ${\cal E}$ is the induced subset of edges of ${\cal E}[{\cal D}_r(t)]$. Let ${\cal N}_{\rm src}[{\cal D}_r(t)]$ be the set of source nodes (i.e., nodes with at least one in-coming directed edge).

For such a subgraph ${\cal H}_r$, we define $W({\cal H}_r)$ as the number of vehicles that can be satisfied by performing index coding on ${\cal H}_r$. We aim to find the best subgraph ${\cal H}_r$ that maximizes $W({\cal H}_r )$ subject to the capacity constraint ${\sf 1J\mbox{-}IdxCd}^{\sf s}[{\cal H}_r] \le {\rm C}^{\downarrow}_r$. Since the number of roads connecting a junction is small, this process can be performed efficiently. For each demand packet that cannot be accommodated by local broadcast, the scheduler will download it via the cellular network. 

First, a greedy online opportunistic scheduling scheme is presented in Algorithm ${\sf ONLSchd}$, which schedules the local broadcast transmissions at RSU based on the arrival of vehicles in an online manner. Static map data will be scheduled before dynamic map data. If there is insufficient capacity at the RSU fog nodes, then the scheduler will download the remaining map data via the cellular network.

The greedy offline opportunistic scheduling scheme is presented in Algorithm ${\sf OFLSchd}$. At each RSU fog node, optimal single-junction index coding is employed, considering all autonomous vehicles that approach the junction at the current time. If there is any spare capacity at RSUs, the scheduler will download the undelivered static map data in advance at any RSU fog nodes with spare capacity. Finally, undelivered map data will be downloaded via the cellular network.

\begin{algorithm}
\caption{\small ${\sf ONLSchd}[{\cal D}_r(t_{\sf now})_{r \in {\cal R}}]$}
{\scriptsize
\begin{algorithmic}[1] 

\For {$r \in {\cal R}$}
\State ${\sf c}^{\downarrow}_r(t_{\sf now}) \leftarrow  {\rm C}^{\downarrow}_r$ \newline \phantom{99} \Comment{\textcolor{black}{{\bf Initialize RSU $r$ download capacity}}}
\LeftComment{{\em Download static map data via local broadcast by index coding}}
\State ${\cal H}_r \leftarrow  {\rm argmax}_{{\cal H}} W({\cal H} )$ \newline \phantom{99} subject to ${\sf 1J\mbox{-}IdxCd}^{\sf s}[{\cal H}] \le {\sf c}^{\downarrow}_r(t_{\sf now})$ and ${\cal H}$ is subgraph of  ${\cal D}_r(t_{\sf now})$ 
\newline \phantom{99} \Comment{\textcolor{black}{{\bf Get max \# of vehicles whose demands can be satisfied}}}
\State ${\cal I}^{\sf s}_r(t_{\sf now}) \leftarrow  {\sf 1J\mbox{-}IdxCd}^{\sf s}[{\cal H}_r]$ \newline \phantom{99} \Comment{\textcolor{black}{{\bf {Perform Optimal Index Coding on ${\cal H}_r$}}}}
\State${\sf c}^{\downarrow}_r(t_{\sf now}) \leftarrow {\sf c}^{\downarrow}_r(t_{\sf now})  -  |{\cal I}^{\sf s}_r(t_{\sf now})|$ \newline \phantom{99} \Comment{\textcolor{black}{\bf Update RSU $r$ download capacity}} 
\For {$c \in {\cal C}_r(t_{\sf now}), e \in {\sf P}^c \cap {\cal E}_r$}
\If{$m^{\sf s}_e \in {\cal N}_{\rm src}[{\cal H}_r]$}
\State $x^{\sf s}_r(t_{\sf now},e) \leftarrow 1$ \newline \phantom{99} \Comment{\textcolor{black}{\bf Download static map data via local broadcast}}
\Else
\State $y_{\sf s}^c(t_{\sf now},e) \leftarrow 1$  \newline \phantom{99} \Comment{\textcolor{black}{\bf Download undelivered static map data via cellular networks}}
\EndIf
\EndFor

\LeftComment{{\em Download dynamic map data via local broadcast by index coding}}
\State ${\cal H}'_r \leftarrow  {\rm argmax}_{{\cal H}'} W({\cal H}' )$ \newline \phantom{99} subject to ${\sf 1J\mbox{-}IdxCd}^{\sf d}[{\cal H}'] \le {\sf c}^{\downarrow}_r(t_{\sf now})$ and ${\cal H}'$ is subgraph of  ${\cal D}_r(t_{\sf now})$ \newline \phantom{99} \Comment{\textcolor{black}{{\bf Get max \# of vehicles whose demands can be satisfied}}}
\State ${\cal I}^{\sf d}_r(t_{\sf now}) \leftarrow  {\sf 1J\mbox{-}IdxCd}^{\sf d}[{\cal H}'_r]$\newline \phantom{99} \Comment{\textcolor{black}{{\bf {Perform Optimal Index Coding on ${\cal H}'_r$}}}}
\State ${\sf c}^{\downarrow}_r(t_{\sf now}) \leftarrow {\sf c}^{\downarrow}_r(t_{\sf now})  -  |{\cal I}^{\sf d}_r(t_{\sf now})|$ \newline \phantom{99} \Comment{\textcolor{black}{\bf Update RSU $r$ download capacity}}
\For {$c \in {\cal C}_r(t_{\sf now}), e \in {\sf P}^c \cap {\cal E}_r$}
\If{$m^{\sf d}_e \in {\cal N}_{\rm src}[{\cal H}'_r]$}
\State $x^{\sf d}_r(t_{\sf now},e) \leftarrow 1$ \newline \phantom{99} \Comment{\textcolor{black}{\bf Download static map data via local broadcast}}
\Else
\State $y_{\sf d}^c(t_{\sf now},e) \leftarrow 1$ \newline \phantom{99} \Comment{\textcolor{black}{\bf Download undelivered dynamic map data via cellular networks}}  
\EndIf
\EndFor

\EndFor
\end{algorithmic}
}
\label{alg:ONLSchd}
\end{algorithm}

\begin{algorithm}
\caption{\small ${\sf OFLSchd}[{\cal D}_r(t)_{t \in {\cal T}, r \in {\cal R}}]$}
{\scriptsize
\begin{algorithmic}[1] 
\For {$t \in {\cal T}, r \in {\cal R}$}
\State ${\sf c}^{\downarrow}_r(t) \leftarrow  {\rm C}^{\downarrow}_r$ 
\newline \phantom{99} \Comment{\textcolor{black}{{\bf Initialize RSU $r$ download capacity}}}
\LeftComment{{\em Download static map data via local broadcast by index coding}}
\State ${\cal H}_r \leftarrow  {\rm argmax}_{{\cal H}} W({\cal H} )$ \newline \phantom{99} subject to ${\sf 1J\mbox{-}IdxCd}^{\sf s}[{\cal H}] \le {\sf c}^{\downarrow}_r(t_{\sf now})$ and ${\cal H}$ is subgraph of  ${\cal D}_r(t_{\sf now})$
\newline \phantom{99} \Comment{\textcolor{black}{{\bf Get max \# of vehicles whose demands can be satisfied}}}
\State ${\cal I}^{\sf s}_r(t) \leftarrow  {\sf 1J\mbox{-}IdxCd}^{\sf s}[{\cal H}_r]$
\newline \phantom{99} \Comment{\textcolor{black}{{\bf {Perform Optimal Index Coding on ${\cal H}_r$}}}}
\State ${\sf c}^{\downarrow}_r(t) \leftarrow {\sf c}^{\downarrow}_r(t)  -  |{\cal I}^{\sf s}_r(t)|$
\newline \phantom{99} \Comment{\textcolor{black}{\bf Update RSU $r$ download capacity}}
\For {$c \in {\cal C}_r(t), e \in {\sf P}^{c} \cap {\cal E}_r$}
\If{$m^{\sf s}_e \in {\cal N}_{\rm src}[{\cal H}_r]$}
\State $x^{\sf s}_r(t,e) \leftarrow 1$
\newline \phantom{99} \Comment{\textcolor{black}{\bf Download static map data via local broadcast}}
\EndIf
\EndFor

\LeftComment{{\em Download dynamic map data via local broadcast by index coding}}
\State ${\cal H}'_r \leftarrow  {\rm argmax}_{{\cal H}'} W({\cal H}' )$ \newline \phantom{99} subject to ${\sf 1J\mbox{-}IdxCd}^{\sf d}[{\cal H}'] \le {\sf c}^{\downarrow}_r(t_{\sf now})$ and ${\cal H}'$ is subgraph of  ${\cal D}_r(t_{\sf now})$
\newline \phantom{99} \Comment{\textcolor{black}{{\bf Get max \# of vehicles whose demands can be satisfied}}}
\State ${\cal I}^{\sf d}_r(t) \leftarrow  {\sf 1J\mbox{-}IdxCd}^{\sf d}[{\cal H}'_r]$
\newline \phantom{99} \Comment{\textcolor{black}{{\bf {Perform Optimal Index Coding on ${\cal H}'_r$}}}}
\State ${\sf c}^{\downarrow}_r(t) \leftarrow {\sf c}^{\downarrow}_r(t)  -  |{\cal I}^{\sf d}_r(t)|$
\newline \phantom{99} \Comment{\textcolor{black}{\bf Update RSU $r$ download capacity}}
\For {$c \in {\cal C}_r(t), e \in {\sf P}^{c} \cap {\cal E}_r$}
\If{$m^{\sf d}_e \in {\cal N}_{\rm src}[{\cal H}'_r]$}
\State $x^{\sf d}_r(t,e) \leftarrow 1$
\newline \phantom{99} \Comment{\textcolor{black}{\bf Download dynamic map data via local broadcast}}
\EndIf
\EndFor

\EndFor

\LeftComment{{\em Download static map data  via local broadcast in advance, if sufficient capacity}}
\For {$t \in {\cal T}, c \in {\cal C}, r  \in {\sf P}^{c}$}
\If{$\exists e \in {\sf P}^{c}$ and $\exists r \in {\sf P}^{c}$ and ${\sf t}^c_r < {\sf t}^c_e$ and $\prod_{t' \le {\sf t}^c_e} (1-x^{\sf s}_r(t',e))=0$ and ${\sf c}^{\downarrow}_r({\sf t}^c_r) \ge  |m^{\sf s}_e|$}
\State $x^{\sf s}_r({\sf t}^c_r,e) \leftarrow 1$
\newline \phantom{99} \Comment{\textcolor{black}{\bf Download static map data via local broadcast}}  
\State ${\cal I}^{\sf s}_r({\sf t}^c_r)  \leftarrow {\cal I}^{\sf s}_r({\sf t}^c_r) \cup \{m^{\sf s}_e \}$
\newline \phantom{99} \Comment{\textcolor{black}{\bf Advanced static map data to be downloaded}}
\State ${\sf c}^{\downarrow}_r(t) \leftarrow {\sf c}^{\downarrow}_r(t)  -   |m^{\sf s}_e|$
\newline \phantom{99} \Comment{\textcolor{black}{\bf Update RSU $r$ download capacity}}
\EndIf
\EndFor

\LeftComment{{\em Download undelivered map data via cellular networks}}
\For {$t \in {\cal T}, r \in {\cal R}$}
\For {$c \in {\cal C}_r(t), e \in {\sf P}^c \cap {\cal E}_r$}
\If{$x^{\sf s}_r(t,e) \ne 1$}
\State $y_{\sf s}^c(t,e) \leftarrow 1$  
\newline \phantom{99} \Comment{\textcolor{black}{\bf Download undelivered static map data via cellular networks}}  
\EndIf
\If{$x^{\sf d}_r(t,e) \ne 1$}
\State $y_{\sf d}^c(t,e) \leftarrow 1$  
\newline \phantom{99} \Comment{\textcolor{black}{\bf Download undelivered dynamic map data via cellular networks}}  
\EndIf
\EndFor
\EndFor

\end{algorithmic}
}
\label{alg:OFLSchd} 
\end{algorithm}

%% file: Upload.tex
\section{Uploading 3D LIDAR Point Cloud Data} \label{sec:octree}

This section focuses on the discussion of 3D LIDAR point cloud data, and a common representation called Octree. We present differential coding and hashing schemes especially for uploading 3D LIDAR point cloud data.

\subsection{Octree Representation}

\begin{figure}[htb!] 
	\centering 
            \includegraphics[scale=1.0]{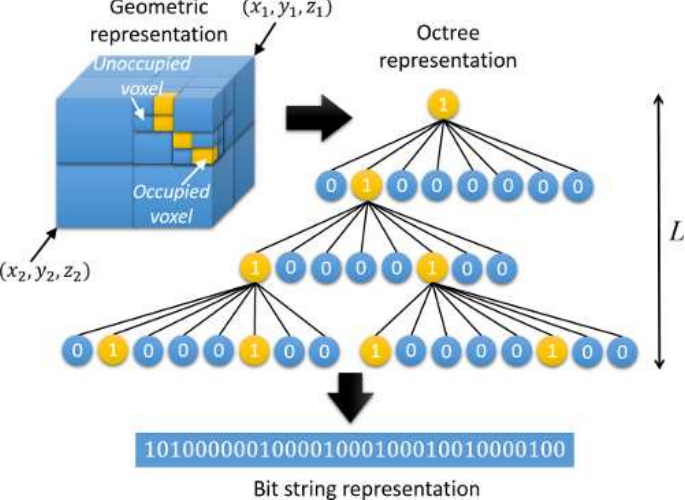} 
        \caption{\small An illustration of Octree representation of 3D point cloud.}\label{fig:octree}
\end{figure}

3D point cloud depicts objects and surfaces as a set of 3D points in the Cartesian coordinate system within a bounded region \cite{Rusu2008}. A common approach to encode 3D point cloud is using {\em Octree}, by which the 3D space is recursively partitioned into 8 cells ({\em voxels}) and a binary number is used to indicate the presence of an object in each cell. See an illustration of Octree representation of 3D point cloud in Fig.~\ref{fig:octree}.

Octree is a tree-based data structure suitable for sparse 3D point data, where each node represents a cell or volume element (voxel).  From the root, it is iteratively divided into eight children until a certain depth or level $L$ is achieved \cite{Kammerl2012} or if there is no more 3D point cloud to be partitioned.  An occupied voxel contains a point or a set of points, and is labeled by `1', otherwise by `0'.  A node labeled by `1' can be further decomposed into eight more child nodes, whereas there is no need to expand a node labeled by `0'. Accordingly, the larger the depth (i.e., higher value of $L$), the higher the resolution of the 3D object.  

Two reference corners for the boundary of region of an Octree are denoted by $(x_1, y_1, z_1)$ and $(x_2, y_2, z_2)$ (see Fig.~\ref{fig:octree}). One can represent an Octree by a bit string representation that encodes its contents by a fixed traversal order in the voxels of each layer. We can apply further coding schemes on the bit string representation. Note that different LIDAR sensors may produce different sets of 3D LIDAR point cloud data on the same objects in the region because of different sensing specifications. But the Octree representations can approximate closely with each other, under a suitable value of $L$. Hence, it is possible to compare different sets of 3D LIDAR point cloud data in Octree representations. 

There are several proposals for point cloud compression \cite{Schnabel2006,Huang2008}. These techniques can be applied to our system, but note that they are mainly for storage and are not optimized for communication systems.

\subsection{Differentiation and Differential Coding} \label{sec:diff}

Autonomous vehicles can identify and upload the necessary dynamic map data to LCD using differentiation. Since dynamic map data is only detectable at the moment of departing from a road segment, the upload transmissions take place immediately through the nearby RSU fog node (in short-range broadcast), whenever possible. Otherwise, cellular network transmissions are employed.

Differentiation is particularly useful for identifying the dynamic components in 3D LIDAR point cloud data. We denote the differentiated data between observed point cloud $x_c(t)$ and reference point cloud $m_e(t-1)$ by:

\begin{equation}
{\sf Diff}_c(t) = \Big( x_c(t) \backslash m_e(t-1) \Big) \cup \Big( m_e(t-1) \backslash x_c(t) \Big)
\end{equation}
where $t = {\sf t}^{c}_{e}$ and $e \in P_c$.

\begin{figure}[htb!] 
	\centering 
            \includegraphics[scale=1.0]{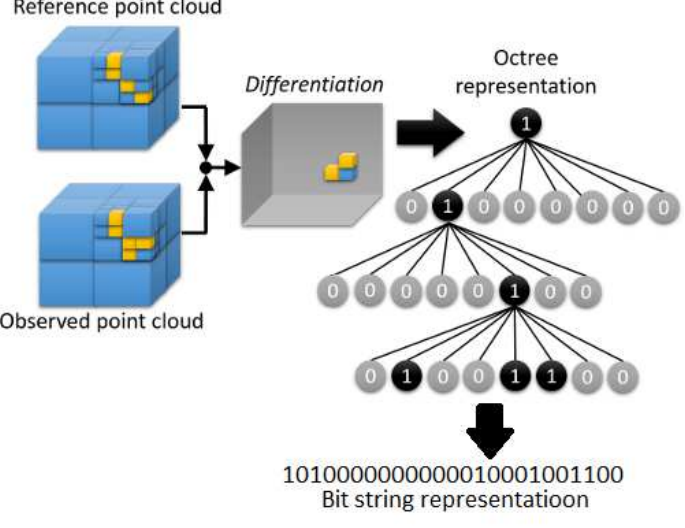} 
        \caption{\small An illustration of differential coding on 3D point cloud.}\label{fig:differentiate}
\end{figure}

To encode the differentiated data, we employ {\em differential coding} on Octree. Octree allows efficient identification of the differences by enumerating the voxels along the tree. Once the differences are identified, we can employ another Octree to encode the differentiated parts. However, the meanings of voxels are now different: `0' means no difference with respect to the reference 3D LIDAR point cloud data, whereas `1' means the binary content in the respective voxel should be flipped. See an illustration in Fig.~\ref{fig:differentiate}.

\subsection{Hashing 3D LIDAR Data} \label{sec:bloom}

Comparison through the hash files associated with 3D LIDAR point cloud data is more efficient than using the whole data set. The hash files should have certain desirable properties. For example, one can compare two hash files to identify which point cloud data consists of more contents (e.g., more observed objects). Second, one can check if the point cloud data contains certain known objects, without looking at the whole data set.   A simple solution is to use a Bloom filter \cite{MMbook}, a compact lossy data structure representing the membership of a set of elements. The basic operations of a Bloom filter involve adding an element to the set and querying the membership of an element.  It does not support element removal, therefore, upon query of an element membership, the Bloom filter output may only result in false positives, which can be minimized through parameter setting. In our system, each vehicle first communicates with RSU and LCD using Bloom filters before uploading the whole perception data. 

{Recall that $x_c(t) = \{p_1, p_2 , ...\}$ is a set of 3D points. Note that each $p_i$ has a unique octary representation, such that each digit in the octary representation represents the order of the respective occupied voxel at each layer in Octree. We denote index `0' to represent the first voxel. For example, the four 3D points in the Octree of Fig.~\ref{fig:octree} can be represented in octary representation as $\{101, 105, 150, 155\}$.  Next, we map each point in octary representation by a set of $K$ binary hash functions: $f_k(p_i) \mapsto \{0, 1\}$, where $k = 1,...,K$. Let $f_k(x_c(t)) = f_k(p_1)\vee f_k(p_2)\vee ...$ be the bitwise disjunction of all the points in $x_c(t) = \{p_1, p_2 , ...\}$. The $K$ output bits $\big(f_k(x_c(t))\big)_{k=1}^K$ will be a Bloom filter for $x_c(t)$, denoted by ${\sf BF}(x_c(t))$. Bloom filters have some desirable properties. If a 3D point cloud has more contents, then its Bloom filter contains more 1's. One can check if a 3D point cloud contains a set of known 3D points, by checking if its Bloom filter contains the corresponding hash values.}

%% file: Testbed.tex
\section{{Robotic Testbed Evaluation}} \label{sec:testbed}
\begin{figure*}[htb!]
	\centering
	\includegraphics[scale=0.7]{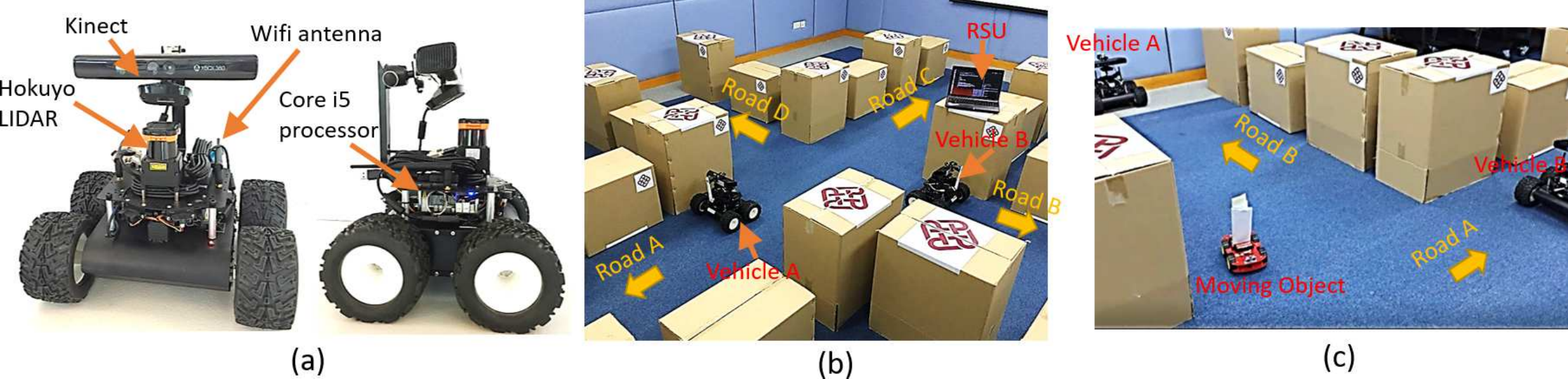} \vspace{-5pt}
	\caption{\small (a) Robotic vehicles in (b) a single 4-way road junction scenario. (c) A moving object is introduced on Road B}\label{fig:Wifibot_City} \vspace{-5pt}
\end{figure*}

\begin{figure*}[htb!]
	\centering
	\includegraphics[scale=0.7, angle=0]{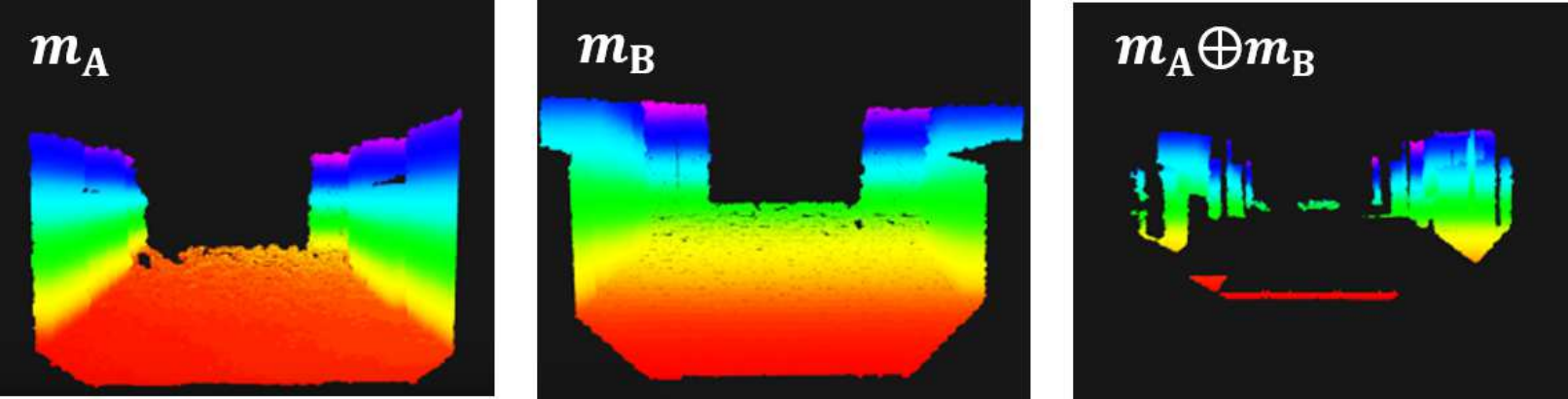} 
	\caption
	{\small The 3D point cloud road map data captured by (a) car A, (b) car B and (c) the XOR-ed result of maps A and B.}\label{fig:3DPCDRepresentations}
\end{figure*}

{We implemented the single junction scenario and evaluated our proposed system in a practical testbed.  In this set-up, as depicted in Fig.~\ref{fig:Wifibot_City}, two cases are studied:}
\begin{itemize}

	\item {{\em Scenario 1}: Car A on Road A intends to turn into Road B with Car B.  There is no time-sensitive data.}
	
	\item {{\em Scenario 2}:  Similar to scenario 1, but there is a moving object in front of Car B on Road B.}
	
\end{itemize}

{The robotic vehicles used in the testbed are shown in  Fig.~\ref{fig:Wifibot_City}(a). They represent the intelligent vehicles equipped with suite of sensors, including a Kinect camera {and LIDAR} for capturing its environment's 3D point cloud data and proximity sensors for collision detection.  The experimental set-up is shown in Fig.~\ref{fig:Wifibot_City}(b). The cardboard boxes represent buildings (see Fig.~\ref{fig:Wifibot_City}).  The 3D point cloud data is compressed to a 1 cm$^3$ resolution to achieve at least a 60$\%$ compression rate before being transmitted to the RSU fog node.  Such Octree resolution offers a significant compression rate while maintaining an accurate representation of the sensed environment.  The RSU fog node and robotic vehicles exchange information by using the IEEE 802.11 standard (WiFi).}

{In Scenario 1, at every five seconds, both vehicles captured their respective environment in form of 3D point cloud data, and performed Octree compression. The data are then transmitted to the RSU fog node along with their requests of road map data.  Upon reception, the RSU fog node performs the {encoding $m_A$ $\oplus$ $m_B$, where $m_i$ is the map data for road segment $i$}, and broadcasts the encoded packets.  The 3D point cloud data perceived by individual vehicle and the corresponding  encoded {3D point cloud data} are shown in Fig.~\ref{fig:3DPCDRepresentations}.  After receiving the encoded packets ($m_A$ $\oplus$ $m_B$), car A decodes it via ($m_A$ $\oplus$ $m_B$) $\oplus$ $m_A$ to obtain its desired information regarding road segment B.  Car B does the same to acquire information regarding road segment A.  Since both road segments have no obstacles detected, each vehicle immediately turns to its desired road without the need of reducing its speed.}

{In Scenario 2, a small programmable mobile robot is added in front of car B to introduce a dynamic object to the environment.  Such set-up is depicted in Fig.~\ref{fig:Wifibot_City}(b).  In order to detect obstacles that present on the road, we integrated a map filter for object extraction after the decoding process, and we search for the blocked information on the ground to determine the location of the object. Fig.~\ref{fig:Wifibot_City}(c) illustrates the detected dynamic data by car A after map filtering and object detection. From the gathered information, car A reduces its speed and waits until the small robot moves past the junction before turning into Road B.} 

{In summary, our robotic testbed manages to achieve cooperative autonomous driving through transmitting road map data between the two robot vehicles.  It also experimentally demonstrates that an efficient 3D road map data dissemination based on the proposed index coding scheme is feasible in practice, especially when dealing with moving dynamic objects on road.}

%% file: Sim.tex
\begin{figure*}[htb!]
		\centering
            \includegraphics[scale=0.6]{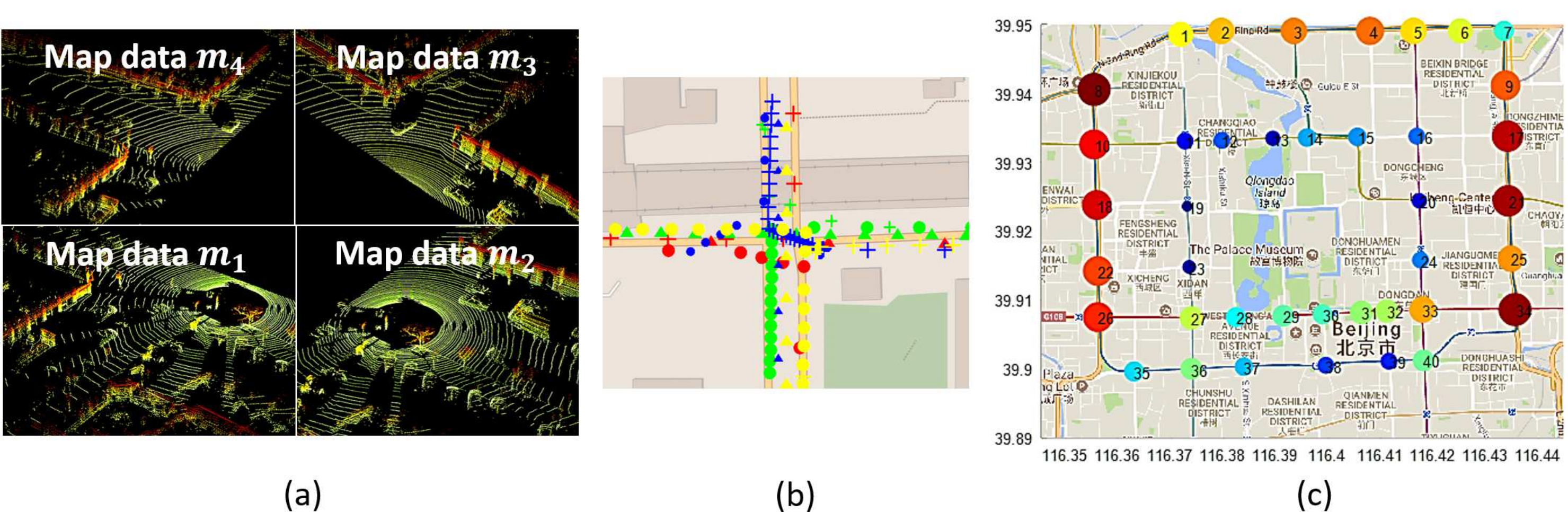} \vspace{-5pt}
        \caption{\small (a) Partitioned 3D point cloud map data of a real-world junction in Fig.~\ref{fig:pointcloud}.  (b) Empirical GPS mobility traces for a particular single junction in Beijing City. (c) Average mobility patterns of 40 selected junctions in Beijing City.}\label{fig:juncdata} \vspace{-5pt}
\end{figure*}

\section{Simulation Studies} \label{sec:sim}

{In the previous section, we have demonstrated the feasibility of employing index coding in the dissemination of road segment data to nearby vehicles at a road junction.  In this section, we present further evaluation of our proposed system by simulation studies using real-world 3D point cloud data of city streets and  GPS mobility traces of vehicles.}  We consider both scenarios of single and multiple road junctions for analyzing the effectiveness of the proposed schemes.

\subsection{Local Broadcast by Index Coding for Single Junction} \label{sec:SingJunc}
First, we consider the single-junction scenario.  The simulation set-up is described as follows.

\medskip

\subsubsection{Simulation Set-up}

To study the performance on realistic 3D point cloud data, we consider the 3D point cloud static map data of a real-world junction depicted in Fig.~\ref{fig:juncdata} (a), which is obtained by Ford Research campus in downtown Dearborn, Michigan \cite{ford}. It is partitioned into four separate views as perceived by the vehicles in each road segment connecting to the junction. In the dissemination process, the 3D point cloud data is compressed using Octree compression \cite{Schnabel2006}.  The sizes of compressed 3D point cloud data packets of each road segment and binary-coded packets are shown in Table \ref{Tab:MapDataAtt}.

\begin{table}[htb!]
	\caption{\small Sizes of compressed 3D point cloud data for the static map data shown in Fig. 9 (a).} \label{Tab:MapDataAtt} 
	\centering
	\begin{tabular}{c|c|c}
		\hline\hline
		\centering Map Data	 &  Number of Points & Data Size (MB) \\
		\hline
		\centering $m_1$ & 104,255 & 5.838 \\		

		\centering $m_2$ & 95,537 & 5.254 \\

		\centering $m_3$ & 69,200 & 2.763 \\

		\centering $m_4$ & 73,168 & 3.184 \\

		\centering $m_1$$\oplus$$m_2$ & 63,607 & 2.126 \\

		\centering $m_1$$\oplus$$m_3$ & 65,920 & 2.812 \\

		\centering $m_1$$\oplus$$m_4$ & 61,738 & 2.630 \\

		\centering $m_2$$\oplus$$m_3$ & 67,806 & 2.631 \\

		\centering $m_2$$\oplus$$m_4$ & 66,072 & 2.363 \\

		\centering $m_3$$\oplus$$m_4$ & 64,025 & 2.126 \\		
		\hline\hline
	\end{tabular}
\end{table}	

To incorporate realistic vehicle mobility patterns, we consider the dataset of Beijing taxi GPS mobility traces \cite{Li2013} to simulate the mobility traces of autonomous vehicles at a junction.  The Beijing taxi dataset contains seven days of GPS mobility traces (including longitude and latitude positions), timestamps of recorded positions, and vehicle IDs of 28,590 taxis traveling in Beijing City.  Beijing City resembles a grid network geographically, consisting of mostly four-way junctions. In particular, we consider the junction between the {\em East 3rd Ring Road Middle} and {\em Jianguo Road}, as shown in Fig.~\ref{fig:juncdata}(b).  There are 8,663 taxis on average traversing it daily.  Fig.~\ref{fig:juncdata}(b) depicts the empirical GPS mobility traces of 12 taxis. We assume that the RSU fog node is deployed near the junction center with a transmission range of 200 meters.

\medskip

\subsubsection{Evaluation of Download Operations}
	
To perform the download operations, a RSU fog node $r$ first scans the nearby vehicles in every sampling time $T_{\rm S}$. Once the vehicles reach within the proximity of $r$, it determines the vehicles' map data demands and constructs the demand graph ${\cal D}_r$. Next, RSU fog node $r$ applies {\sf 1J-IdxCd}$[{\cal D}_r]$ to perform local broadcast based on index coding. 

To study the performance of {\sf 1J-IdxCd}$[{\cal D}_r]$, we consider two benchmarks: 
\begin{enumerate}

\item Random Broadcast ({\sf Rand}): It broadcasts all source packets in a random fashion. 

\item Index Coding with Prior Information ({\sf 1J-IdxCd-PI}): It explores the scenario that some vehicles may have extra prior knowledge of a certain road segment. For example, a particular road segment is popular among all vehicles. The map data is likely to be pre-downloaded to the vehicles in advance.  

\end{enumerate}

\begin{figure}[htb!]
	\centering
	\includegraphics[scale=0.6, angle=0]{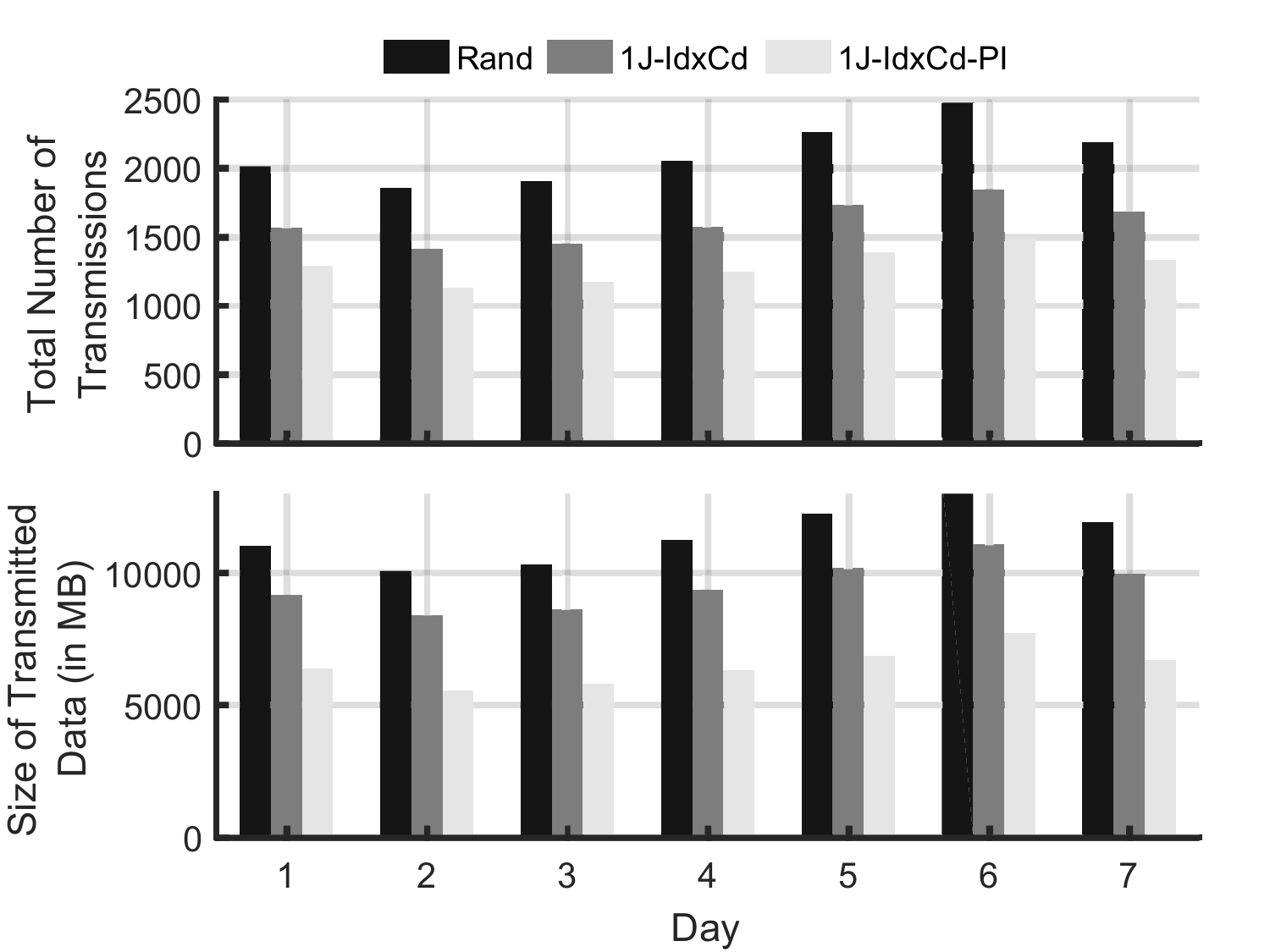} 
	\caption{\small Daily total number of transmissions and sizes of transmitted packets for {\sf Rand}, {\sf 1J-IdxCd}, {\sf 1J-IdxCd-PI}.}\label{fig:WeekTrans_3scheme}
\end{figure}

The evaluation results are depicted in Fig.~\ref{fig:WeekTrans_3scheme}, which shows the daily total number of transmissions and sizes of transmitted packets for {\sf Rand} and {\sf 1J-IdxCd} for seven days based on GPS mobility traces. The sizes of each transmitted packets are set according to Table~\ref{Tab:MapDataAtt}.  

It is observed that {\sf 1J-IdxCd} can effectively reduce the total number of transmissions by around 500 transmissions less when compared to the benchmark {\sf Rand}.  For downloading  static data, the benchmark requires a number of 7.75 transmissions on average to satisfy all vehicles' demands as compared to {\sf 1J-IdxCd} that requires only 5.94 transmissions on average.  The average daily sizes of transmitted data for random transmission is  12.18 GB while that for {\sf 1J-IdxCd} is only 10.24 GB.  {\sf 1J-IdxCd} transmits 6.00 MB on average within a period of $T_{\rm S}$, while {\sf Rand} transmits 5.46 MB on average.  Overall, employing {\sf 1J-IdxCd} enables higher data rate with the fewest number of transmissions.

Next, we evaluate the effectiveness with extra prior information. {\sf 1J-IdxCd-PI} considers extra prior information for road segment 2.  In this case, {\sf 1J-IdxCd-PI} preforms like {\sf 1J-IdxCd}, but it assumes every vehicle already has map data of road segment 2, and performs index coding incorporating such prior information. We observe that the availability of extra prior information considerably reduces the number of transmissions, and thus the required bandwidth, transmitting 5.01 MB on average by 4.49 transmissions.

\subsection{Applying Index Coding to Multiple Junctions}

After evaluating the performance of single-junction index coding, we consider index coding for multiple junctions. 

\medskip
\subsubsection{Simulation Set-up}

We selected 40 junctions in Beijing, as depicted in Fig.~\ref{fig:juncdata}(c), and use the corresponding GPS mobility traces of taxis traversing these junctions to simulate the mobility patterns. The simulation parameters are listed in Table~\ref{Tab:MJSimParam}.  In Fig.~\ref{fig:juncdata}(c), we visualize the average mobility patterns of these 40 junctions by circles of different sizes. The bigger the circle, the more number of taxis traversed the respective junction.   

\begin{table}[htb!]
	\centering
	\caption{Simulation Parameters for Multiple Junctions in Fig.~\ref{fig:juncdata}(c).}
	\label{Tab:MJSimParam}
	\vspace{-5pt}
	\begin{tabular}{l|c}
		\hline\hline
		\textbf{\centering Simulation Attribute/Parameter}        & \textbf{Value}        \\
		\hline
		 Total area (in $\approx$ km$^2$)   & 50                    \\
		Number of observed days                      & 7                     \\
		Number of road segments per junction      & {4} \\
		Number of RSUs        & 40                    \\
		RSU transmission range (meters)                   & 200                   \\
		Total number of taxis                           & 24,845                 \\
		Daily average number of taxi trips        & 79,012                 \\
		Hourly average number of taxis in each junction & 466                   \\
		Total number of recorded time each day (hrs)     & 24                    \\
		Sampling time of GPS traces  (mins)                & 2      \\
		\hline\hline              
	\end{tabular}
\end{table}

\subsubsection{Evaluation Results}

\begin{figure}[htb!]
	\centering
	\includegraphics[scale=0.6, angle=0]{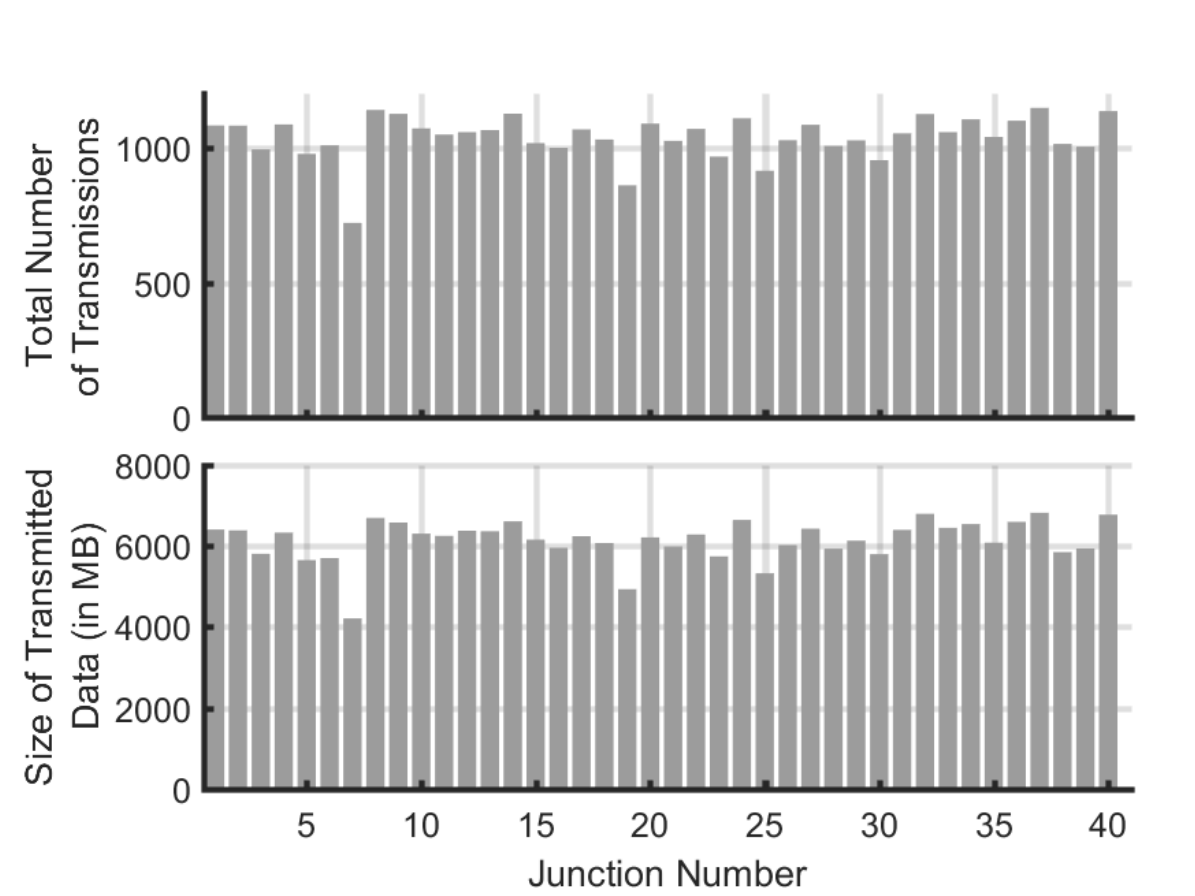} \vspace{-10pt}
	\caption
	{\small Average total number of transmissions and sizes of transmitted data for each of the 40 RSU fog nodes situated in Beijing.}\label{fig:RSUs40Results}
\end{figure}

Fig.~\ref{fig:RSUs40Results} depicts the average number of transmissions and sizes of transmitted data when {\sf 1J-IdxCd} is applied independently at the 40 junctions. Note that the number of visits is not directly proportional to the average number of transmissions.  In particular, RSU fog nodes 19 and 24 have relatively low volume of visits, whereas RSU fog nodes 9 and 25 have a relatively high number.  However, RSU fog node 19 has fewer number of transmissions than RSU fog node 24. This is because the vehicles in RSU 19 arrive more regularly than those at RSU 24, hence more significant performance gain can be found in terms of the number of transmissions and sizes of transmitted data by {\sf 1J-IdxCd}. A similar phenomenon is observed at the high-volume RSU fog node 25.

\begin{figure}[htb!]
	\centering
	\includegraphics[scale=0.7, angle=0]{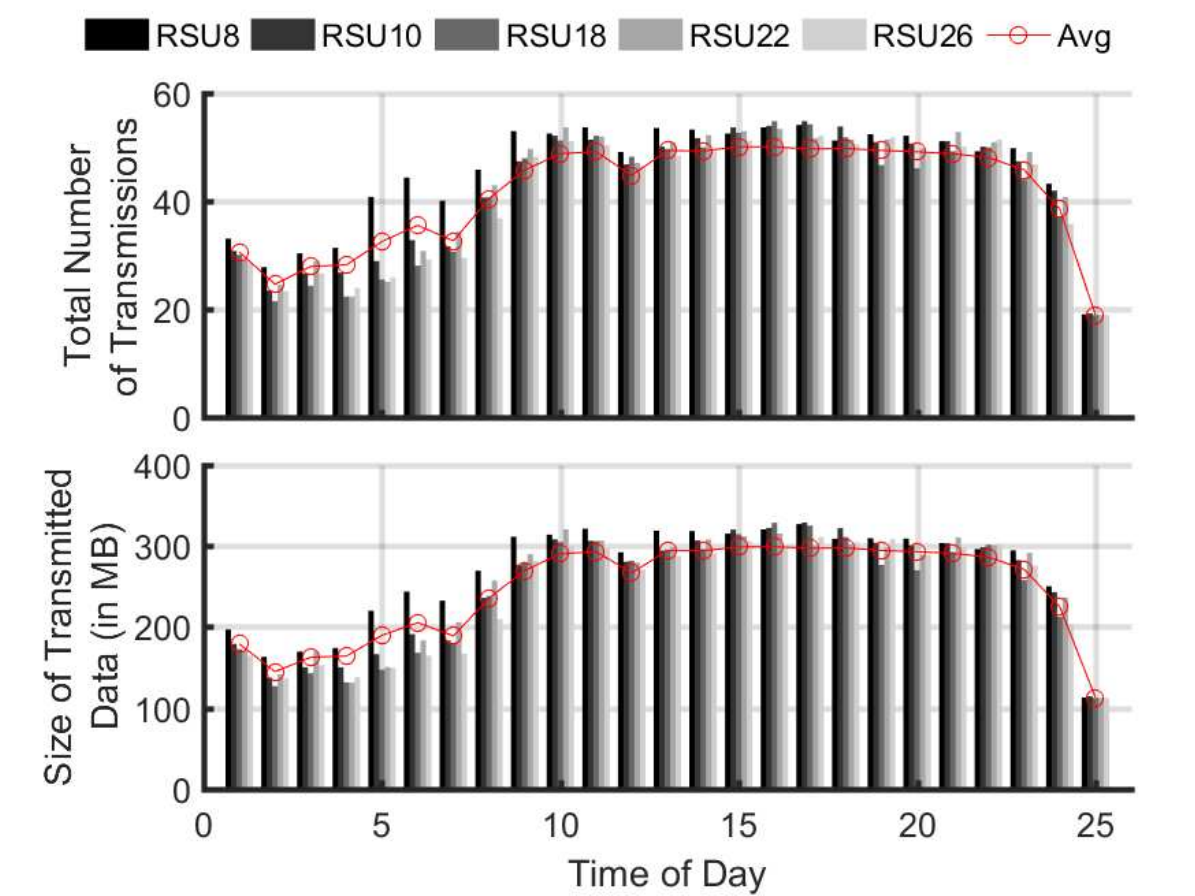} \vspace{-10pt}
	\caption
	{\small Hourly average total number of transmissions and sizes of transmitted data for each RSU fog nodes located on {\em West 2nd Ring Road}.}\label{fig:RSUs40Hourly}
\end{figure}

Fig.~\ref{fig:RSUs40Hourly} shows the RSU fog nodes located on {\em West 2nd Ring Road} (i.e., RSUs 8, 10, 18, 22, 26) and depicts the hourly performance of each RSU. The average volume of visits through these road sections are similar.  The curve labeled by `Avg' indicates  the average value of the 40 RSU fog nodes over a day.  We observe that the information dissemination by the RSU fog nodes increased starting from 08:00h, because the peak traffic hours occur at 08:00h. From midnight to 06:00h, the traffic is relatively low.

\subsection{Scheduling over Multiple Junctions}

In this section, we evaluate the performance of the proposed scheduling schemes over the selected 40 junctions. We employ both the online ({\sf ONLSchd}) and offline ({\sf OFLSchd}) opportunistic scheduling schemes for disseminating map data to vehicles.  

\medskip

\subsubsection{Applying the Opportunistic Scheduling}

\begin{figure}[htb!]
	\centering
	\includegraphics[scale=0.55]{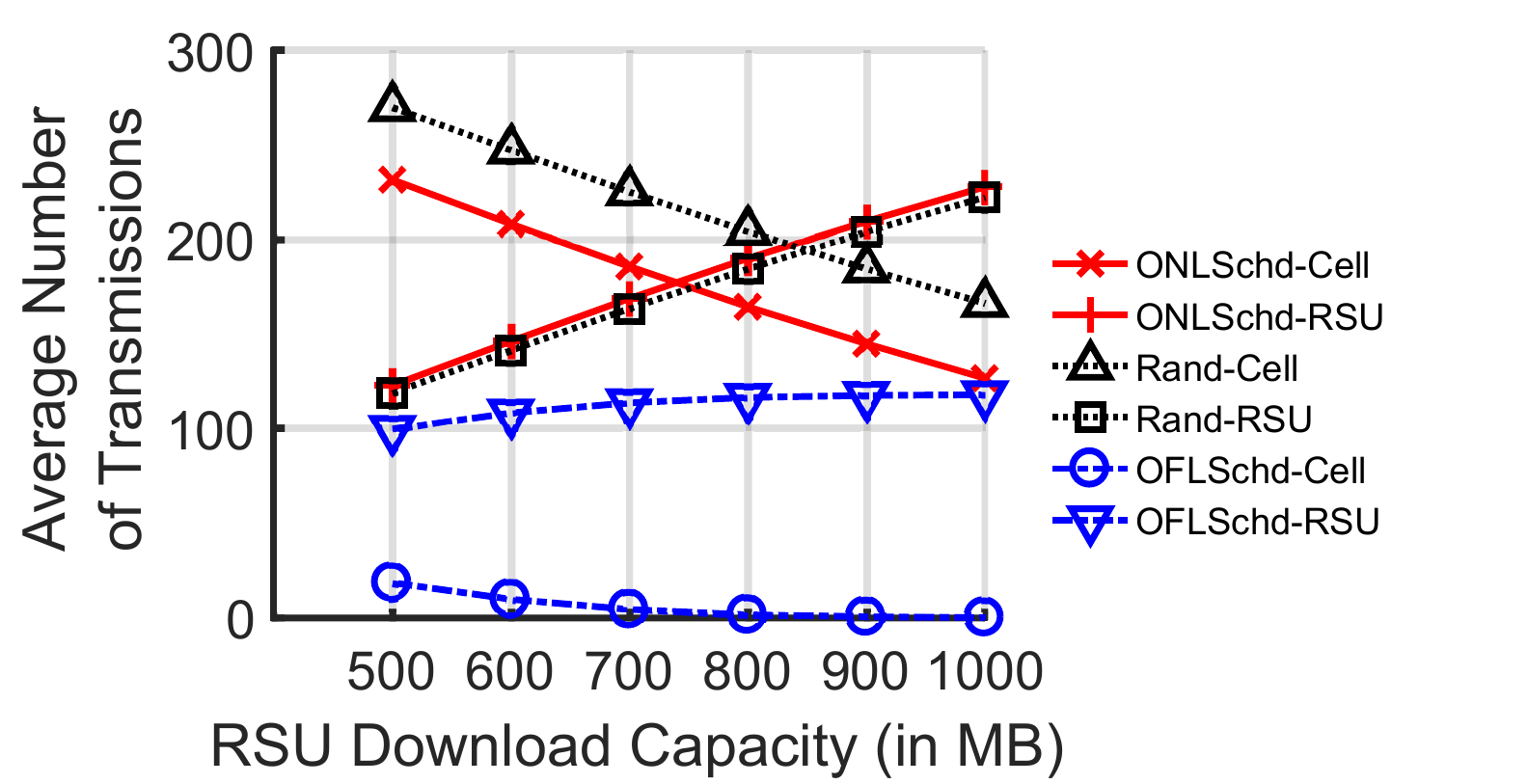} \vspace{-5pt}
	\caption{\small Performance of various scheduling schemes against download capacity (${\rm C}^{\downarrow}$).}\label{fig:GreedySchdResults}
\end{figure}

We assign each RSU fog node the same download capacity (${\rm C}^{\downarrow}$). The performance of various scheduling schemes are shown in Fig.~\ref{fig:GreedySchdResults}, under various download capacity (${\rm C}^{\downarrow}$) based on the GPS mobility traces of taxis traversing the 40 junctions in Beijing in Fig.~\ref{fig:juncdata}(c).

We observe that the RSU fog nodes reach the download capacity at a much faster rate by  {\sf Rand}, because {\sf Rand} broadcasts a large amount of data (equal to the sum of all source packets of map data in a trip per vehicle), as compared to the opportunistic scheduling schemes per sampling time. This leads to heavier load on the cellular network when {\sf Rand} is used in scenarios with low download capacity (${\rm C}^{\downarrow} \leq$ 700 MB).  Although {\sf ONLSchd} reduces the required cellular transmissions, it is still evident that it exhibits the same effect during low download capacity, i.e., more cellular transmissions than local broadcast transmissions.  For a given set of map data, increasing the local broadcast download capacity can reduce the need for cellular network transmissions.  This presents a design trade-off for the network administrator to balance the loads between local broadcast and  cellular unicast.

Among the three schemes, {\sf OFLSchd} employs considerably less cellular network bandwidth even when the local broadcast download capacity is low.  It relies almost totally on local broadcast transmissions as the download capacity is over 700 MB.  This is because all of the vehicles' trip plans are known in advance, thus, enabling the RSU fog nodes to schedule map data dissemination more efficiently, {and less rely on the cellular download of road map data.}

\begin{figure}[htb!]
	\centering
	\includegraphics[scale=0.5]{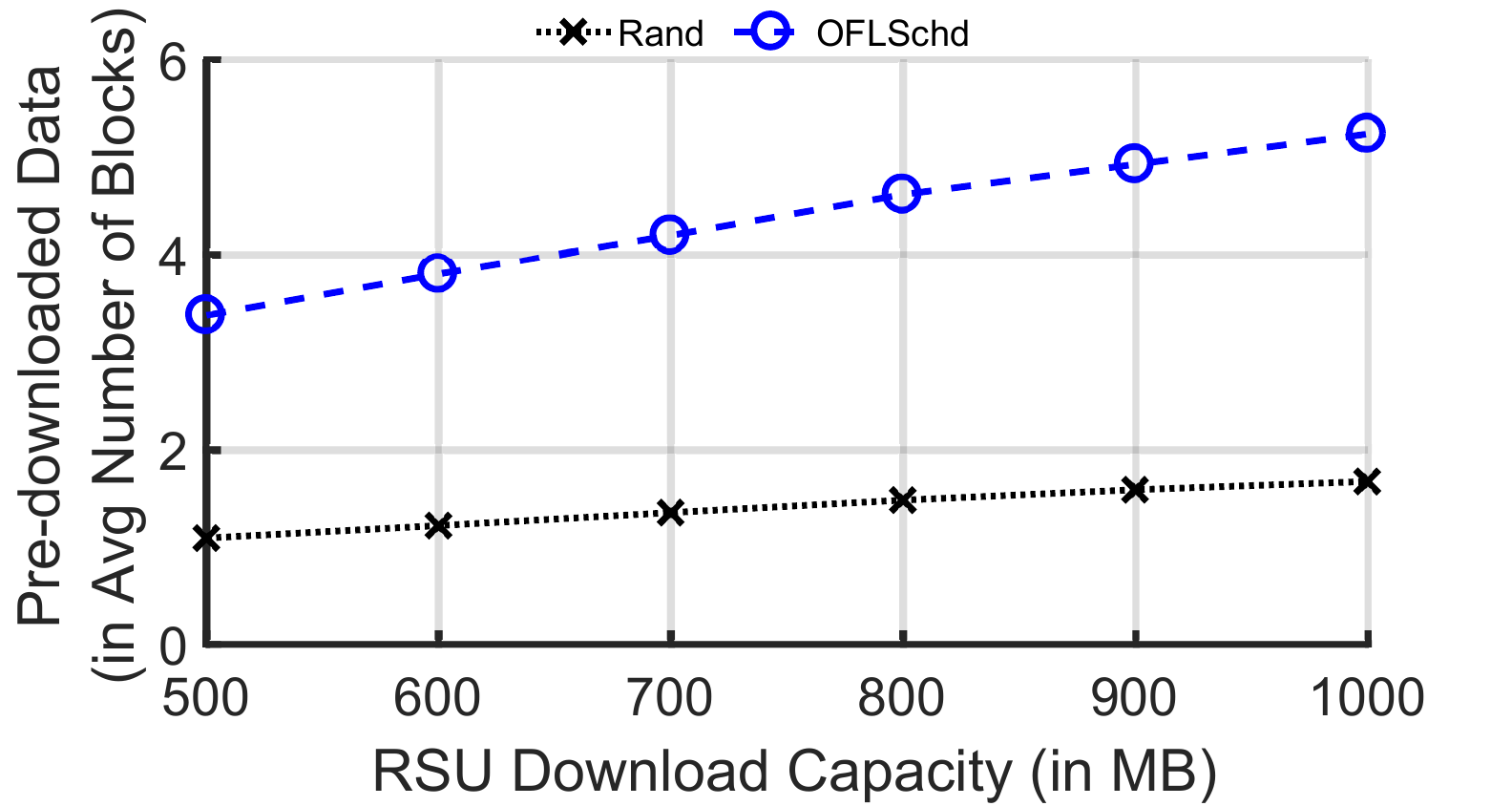} \vspace{-5pt}
	\caption
	{\small Comparing the {\sf OFLSchd} and {\sf Rand} methods on the distance of pre-downloaded remote data to requesting vehicles.}\label{Fig:PreDown}
\end{figure}

Fig.~\ref{Fig:PreDown} shows the distance (in terms of the number of blocks away) of pre-downloaded remote data to a requesting vehicle at an RSU fog node.  Since {\sf OFLSchd} knows the planned trips of the vehicles, it reduces the number of transmissions required to satisfy all vehicles, while providing pre-downloaded remote data up to five blocks away (when the RSU fog node download capacity is $\geq$ 900 MB).  For example, RSU fog node 14 can transmit road segment data from remote RSU fog nodes such as 1, 10, 17, 32, etc. Such amount of advanced data will allow a vehicle to update its planned trip and alter its route if necessary.  On the other hand, {\sf Rand} can only deliver data up to an average of 1.66 RSU blocks away from the requesting RSU fog node. 

\medskip
\subsubsection{Meeting Frequency of Vehicles}

Theorem~\ref{thm:multijunc} shows that applying Algorithm {\sf 1J-IdxCd} independently at each junction produces an optimal index coding scheme for multiple junctions, under the single meeting assumption. In this section, we empirically examine the meeting frequency of vehicles based on GPS mobility traces. 

We define that a meeting occurs when the taxis' routes (according to GPS traces) are within 200 m of each other.  Given an observation window during the day, these frequencies of meetings are stored in the adjacency matrix ${\bf TM}$, defined below in (\ref{Eq:TM}):
\begin{equation}
{\bf TM} = \begin{bmatrix}
0 & x_{1,2} & \cdots & x_{1,n-1}& x_{1,n} \\ 
x_{2,1} & 0 & \cdots & \vdots  & x_{2,n}\\ 
\vdots & \vdots &\ddots  & \vdots & \vdots \\
x_{n-1,1} & \cdots & \cdots & \ddots & x_{n-1,n} \\ 
x_{n,1} & x_{n,2}  & \cdots & x_{n,n-1} & 0
\end{bmatrix}
\label{Eq:TM}
\end{equation}
where $x_{i,j} \in$ $\{0, 1, 2, ...\}$ is the number of {times that} vehicles $i$ and $j$ met and $x_{i,j} = x_{j,i}$. The observation window is equal to $t_{\rm stop} - t_{\rm start}$.  In this case, $t_{\rm stop} - t_{\rm start}$ = 24 hours, having a 2-min sampling interval.  

\begin{figure}[htb!]
	\centering
	\includegraphics[scale=1.0]{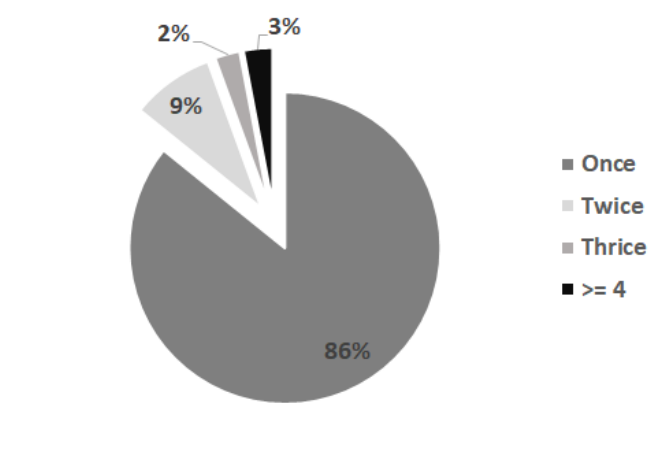} \vspace{-10pt}
	\caption
	{\small  Meeting frequency between any pair of taxis.}\label{fig:FreqofTaxiMeetings}
\end{figure}

 The results are illustrated in Fig.~\ref{fig:FreqofTaxiMeetings}. Any two given taxis from the mobility traces only met each other once 86$\%$ of the time, while the remaining 14$\%$ met more than once.  From Fig.~\ref{fig:FreqofTaxiMeetings}, we know that any pair of taxis met at most once within a moderate time window. This shows limited meetings among vehicles with different trips in a city in practice. Hence, applying Algorithm {\sf 1J-IdxCd} independently at each junction still provides a heuristic to construct a good index coding scheme for a city-wide multiple junction scenario.

\begin{figure}[htb!]
	\centering
	\includegraphics[scale=1.1]{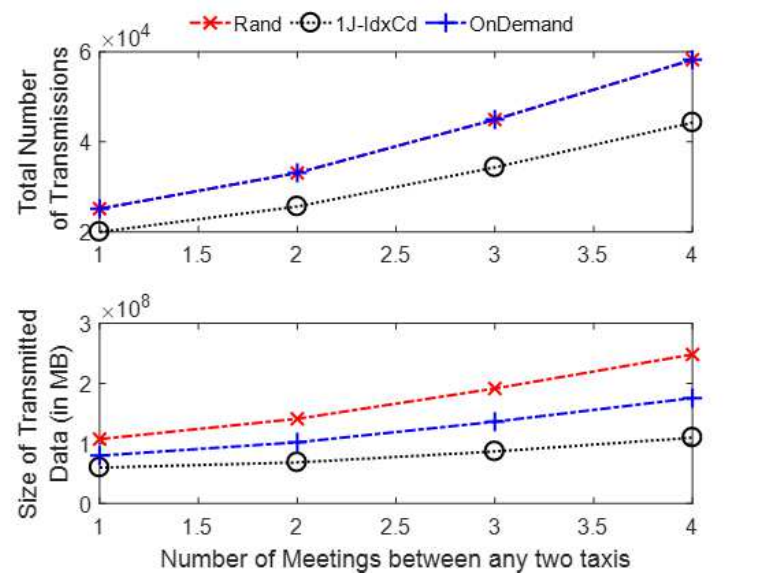} \vspace{-10pt}
	\caption
	{\small Comparison of the three schemes in terms of the total number of transmissions (top) and total sizes of transmitted data (bottom) against the meeting frequencies between any pair of vehicles.}\label{fig:SimFreqofTaxiMeetings}
\end{figure}

Finally, we {study} the scenarios that {the majority of the number} of meetings in the adjacency matrix {\bf TM} is more than one (i.e., $x_{i,j} \ge 2$).  Note that the proposed scheme produces the best performance when $x_{i,j}$ = 1. The simulation results for 3,000 runs are shown in Fig.~\ref{fig:SimFreqofTaxiMeetings}.  We also compare the performance with a benchmark scheme, {\sf OnDemand}  \cite{ali2017efficient}, which transmits the source packets of the most demanded road segment first, and then transmits binary-coded packets of the most demanded road segment data until all demands are satisfied.

We observe that as the number of meetings between any pair of vehicles increases, the total number of transmissions increases in all scheduling schemes (see the top figure of Fig.~\ref{fig:SimFreqofTaxiMeetings}). We note that both {\sf Rand} and {\sf OnDemand}, having equal number of transmissions, increase at a higher rate than {\sf 1J-IdxCd}.  However, {\sf OnDemand} transmits fewer data than {\sf Rand} to satisfy all requesting vehicles.  Overall, considering both performance metrics, {\sf 1J-IdxCd} outperforms the two benchmarks in the multiple junction scenario.
  
\subsection{{Processing Overhead Analysis}} \label{OHAnalysis}

{We analyze in this subsection the processing overheads of the proposed index coding algorithm based on the overall data dissemination delay (including both the processing/encoding delay and transmission delay) from the RSU to the nearby vehicles.  For the {\sf Rand} method, the overall delay only contains the transmission delay, while the {\sf 1J-IdxCd} scheme includes also the processing delay due to the XOR encoding of relevant road map data, which is proportional to the number of encoded packets generated.  To compute for the transmission delay, we assume that the packet size is 1024 bytes and the data rate is 6 Mbps.  On the other hand, the encoding processing delay is assumed to be fixed.  For a given RSU fog node, the overall delay is computed every sampling time $T_S$ = 2 min.}
	
\begin{figure}[htb!]
	\centering
	\includegraphics[scale=1.0]{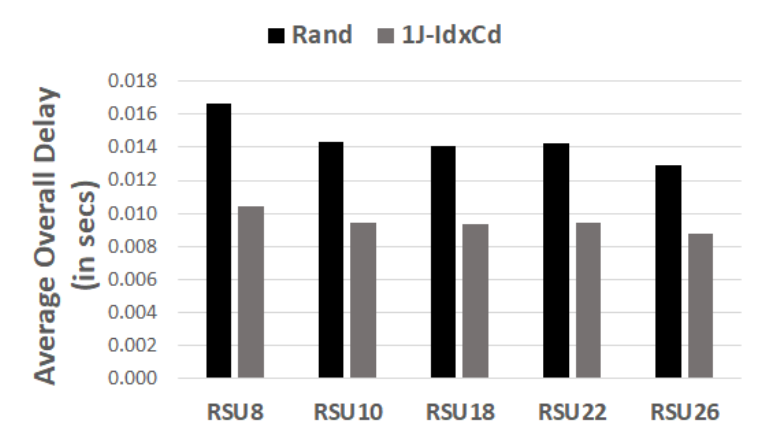} \vspace{-10pt}
	\caption
	{\small Average overall delay of each RSU fog node under the two transmission schemes.}\label{fig:RSUOverallDelay}
\end{figure}

{Given a processing delay of 1 ms, Fig. \ref{fig:RSUOverallDelay} illustrates the overall delay averaged over seven days for RSUs 8, 10, 18, 22, and 26.  {These five RSU fog nodes have a daily average of 8,100 taxis passing through, and there are 11--12 taxis connected to each RSU per $T_S$ on average.}  We can observe that even if there is an additional processing time introduced by the {\sf 1J-IdxCd} method, its daily average overall delay is still less than that of the {\sf Rand} method by about 34\%.  This is because the {\sf 1J-IdxCd} scheme has a much shorter transmission delay than the {\sf Rand} method by reducing the total number of required packets and the number of road segments at each intersection is limited. }

%% file: Conclusion.tex
\section{{Conclusion}} \label{sec:disc}

In this paper, we have presented an efficient information dissemination system of 3D point cloud road map data (3D-MADS) for intelligent vehicles and roadside infrastructure integrated in a vehicular fog computing architecture. Our system aims to minimize the amount of cellular network unicast while maximizing the utility of short-range local broadcast transmissions by implementing fog-based opportunistic schedulers. We have also optimized the performance of 3D point cloud data dissemination and update by utilizing techniques such as index coding at roadside unit fog nodes and hashing of 3D point cloud data at vehicular nodes. The overall system was validated with empirical mobility traces, 3D LIDAR data, and an experimental multi-robotic testbed.